\documentclass[12pt,preprint]{aastex}

\shorttitle{Evolving Starburst Model for Neutral Gas Media}
\shortauthors{Yao}

\begin{document}

\title{Evolving Starburst Modeling of FIR/sub-mm/mm Line Emission. II. Application to M 82}

\author{Lihong Yao\altaffilmark{}}
\affil{Department of Astronomy and Astrophysics, University of Toronto,
    Toronto, ON M5S 3H8, Canada}
\email{yao@astro.utoronto.ca}

\author{}
\affil{Submitted to ApJ on 15 March 2009; Accepted on 3 September 2009}

\begin{abstract}
We present starburst models for far-infrared/sub-millimeter/millimeter
(FIR/sub-mm/mm) line emission of molecular and atomic gas in an 
evolving starburst region, which is treated as an ensemble of 
non-interacting hot bubbles which drive spherical shells of 
swept-up gas into a surrounding uniform gas medium. These bubbles 
and shells are driven by stellar winds and supernovae within massive 
star clusters formed during an instantaneous starburst. The 
underlying stellar radiation from the evolving clusters affects 
the properties and structure of photodissociation regions (PDRs) 
in the shells, and hence the spectral energy distributions (SEDs) 
of the molecular and atomic line emission from these swept-up 
shells and the associated parent giant molecular clouds (GMCs) 
contains a signature of the stage of evolution of the starburst.

The physical and chemical properties of the shells and their 
structure are computed using a a simple well known similarity 
solution for the shell expansion, a stellar population synthesis 
code, and a time-dependent PDR chemistry model. The SEDs for 
several molecular and atomic lines ($^{12}$CO and its isotope 
$^{13}$CO, HCN, HCO$^+$, C, O, and C$^+$) are computed using 
a non-local thermodynamic equilibrium (non-LTE) line radiative 
transfer model.

By comparing our models with the available observed data of 
nearby infrared bright galaxies, especially M 82, we constrain 
the models and in the case of M 82, we provide estimates for 
the ages (5 - 6 Myr, 10 Myr) of recent starburst activity. We also 
derive a total H$_2$ gas mass of $\sim$ 2 - 3.4 $\times$ 10$^8$ M$_{\odot}$ 
for the observed regions of the central 1 kpc starburst disk of M 82. 

\end{abstract}


\section{Introduction} \label{intro}

Knowledge of the physical properties and evolution of the gas and
dust content in the interstellar medium (ISM) of starburst galaxies
is essential for understanding the cause and temporal evolution of
star-forming activity. In particular, studies of such galaxies in
the nearby universe are essential as a step in understanding the
role of the starburst phenomenon in the cosmic evolution of galaxies.
To constrain theories of how the ISM evolves, one needs to investigate
both individual galaxies and large statistical samples of data at
multiple wavelengths. Especially, with the available data for the
dust component, studying the gas in the co-space ISM becomes more
interesting and important.

The number of multiwavelength observations of starburst galaxies 
throughout the cosmic-scale has increased dramatically due to the 
significant improvement in the sensitivity and resolution of telescopes.
These observations provide an essential basis for starburst modeling,
and such modeling provides systematic predictions of the properties 
of the ISM in idealized models of starburst galaxies for comparison 
with these observations. For example, M 82, an irregular starburst 
galaxy ({\it I0}) in the nearby universe has an inclination of 
$81^{\circ}$. The infrared luminosity of M 82 arises mostly from 
the central 1 kpc region, which has a stellar bar structure and 
a complex system of clumps and filaments \citep{las63, bec78, lok87}. 
The formation mechanism of this complex system and the evolutionary 
scheme in M 82 remain under debate \citep[e.g.][]{vis74, cak91, sal96, wil00}. 
Most recently, we conducted an ideal case study of an expanding shell 
model in M 82, and suggested that the circumnuclear rings seen in 
this galaxy may possibly be a consequence of the evolution of 
swept-up gas caused by starbursts that occurred in the center 
$\sim$ 100 Myr ago (Yao et al 2006, hereafter Paper I).

Previously, the age estimates of the starburst in M 82 have been 
presented by many authors. Yun et al. (1993) compared the disk HI 
with optical maps, and found a large amount of gas being channeled 
into the core of the galaxy over the last 200 Myr due to the tidal 
encounter with its large spiral neighbor galaxy M 81. Ages derived 
from super star clusters (SSCs) in optical and near-IR images are 
$\sim$ 50 Myr \citep{deg01}, $\sim$ 30 - 100 Myr \citep{rie93, bar08}, 
$\sim$ 4 - 6, and 10 - 30 Myr \citep{for03, smi06, mel05}. The corresponding 
estimates of the average star formation (SF) rate over the 200 Myr 
period is roughly about 10 M$_{\odot}$ yr$^{-1}$. The importance of 
optical and near-IR spectroscopy in studies of dusty star-forming 
galaxies has long been recognized. But studies of young stellar 
populations at these wavelengths remain difficult. The age determinations 
are affected by residual effects due to the age-metallicity degeneracy, 
and age-IMF degeneracy. In addition, the completeness of the sampled 
stellar population is affected by the unavoidable effects of extinction 
in the optical and near-IR.

The star formation history of M 82 has also been studied using mid-infrared 
and far-infrared spectroscopy \citep{col99, efs00, gal03, sak07}. 
At these wavelengths, the fine structure line emissions are relatively 
insensitive to extinction, and hence can provide a unique probe of age 
and star formation history in an infrared-bright, dust-obscured galaxy 
like M 82. Colbert et al. (1999) obtained a burst age of 3 - 5 Myr for 
the central 1 kpc (65 - 85$^{\arcsec}$) region using an instantaneous 
starburst model and a steady-state PDR model. However, their single burst 
model is dominated by the brightest and most recently formed stars (the 
hot spots seen in mid-IR). Efstathiou et al. (2000) presented an evolving 
starburst model for dusty media using state of the art codes for 
calculating the radiative transfer in dust shells, and incorporating 
a model for the composition and size distribution of grains in the ISM. 
Their study concluded that it is possible to relate the observed 
infrared spectrum of dust associated with a starburst to its age 
and its star formation history by following the evolution of an 
ensemble of GMCs of identical mass induced by massive star formation 
in their centers. They show that the burst age for the central 500 pc 
region of M 82 is between 10 and 30 Myr using a model with two 
instantaneous bursts. Their derived ages are supported by near-IR 
spectroscopy and high-resolution imaging of stellar clusters 
\citep{sat97}. Efstathiou et al. (2000) also suggested that far-IR 
surveys may preferentially detect older starbursts than mid-IR 
studies, based on an argument concerning the evolution of the 
luminosity of starbursts observed at different wavelengths. 
It is clear that the estimation of M 82 starburst age has a large 
uncertainty. The questions we ask are

\begin{center}
{\it Does the FIR/sub-mm/mm molecular and atomic line data contains 
a signal of the evolutionary phase of a starburst?}

{\it Can the molecular sub-mm and mm lines provide an alternative
tool for estimating the starburst ages?}  
\end{center}  

This study is motivated by the abundant evidence of giant bubbles and 
shells found in Milky Way, 30 Doradus in the Large Magellanic Cloud 
(LMC), and nearby starburst galaxies \citep{ped03, deg01}, the success 
of using the dusty starburst model to constrain the star formation history 
of observed IRAS starburst galaxies by following an ensemble of GMCs 
\citep{efs00}, and the available multiple transitions in several molecular 
tracers. In a previous paper (Paper I), as a preliminary approach we presented 
a simple model of an expanding supershell surrounding a massive star 
cluster and its expansion at different evolutionary stages that predict 
sub-mm/mm and far-IR emission line intensities from several key molecular 
and atomic constituents in the shell ($^{12}$CO and its isotope $^{13}$CO, 
HCN, HCO$^+$, C, O, and C$^+$). In this paper, we present a series of 
models, called {\it Evolving Starburst Model} (ESbM) for the molecular 
and atomic line emission from an evolving starburst region, which is 
treated as an ensemble of evolving GMCs each of which is centrally 
illuminated by a compact star cluster (SC). The GMCs in our model ensemble 
follow a power-law mass spectrum. By comparing with the available observed 
data on nearby starburst galaxies, we can constrain the models and provide 
better interpretations for the observations. The main goals of 
this paper are (1) to show that it is possible to model the 
FIR/sub-mm/mm line emission of molecular and atomic gas by following 
the evolution of a starburst region, as in certain infrared (IR) 
models; (2) to relate the observed molecular line properties 
of a starburst galaxy to its age, and hence to constrain the 
global star formation history; (3) to understand the formation 
mechanism of the molecular rings in M 82; (4) and finally, to provide 
useful information for the interpretation of future high resolution 
maps of molecular gas on small and large scales in starburst galaxies, 
in order to provide a deeper understanding of the structure, dynamics, 
and evolution of the neutral ISM and its relationship with active star 
formation.

The plan of this paper is as follows. In \S~\ref{esbm} we discuss 
briefly our model assumptions, initial parameters, and model 
procedure. In \S~\ref{ss} we present examples of our modeling results 
for an ensemble of GMCs/shells. In \S~\ref{m82} we present an application 
of the model to derive the age of the starburst and molecular gas swept up 
by the shells in M 82, and to provide new insights into the nature 
and physical state of the ISM in its starburst region. We also discuss 
applications to the supershell surrounding supernova remnant (SNR) 
41.9 + 58. In the conclusions, we summarize the main results of this work.

Through this study, we hope to provide some answers to those 
intriguing questions mentioned above, and to lay a foundation 
for future starburst modeling for neutral gas media. 

\section{Evolving Starburst Model For Gas Media} \label{esbm}

The basic assumptions for our ESbM model are (1) star formation occurs 
within the dense optically thick spherical cloud (GMC) \citep[e.g.][]{gao01}, 
and star formation takes place instantaneously, with the star 
cluster treated as a point source (see Paper I); (2) absorption of 
the starlight from the central cluster is produced by dust 
associated with the gas, assuming a constant gas-to-dust ratio; 
and (3) the gas responding to star formation in a starburst 
galaxy is treated as an ensemble of GMCs with different initial 
masses, each of which responds to massive star formation at its
center. 

Our ESbM model incorporates a standard similarity model for the 
bubble/shell structure around a young star cluster \citep{wea77, mak87}, 
a time-dependent stellar population synthesis model \citep{lei99}, 
a fully time-dependent chemistry model for the PDRs \citep{bel05, rol07}, 
and a non-LTE radiative transfer model for molecular and atomic lines 
\citep{ray01, vza02}. Theoretical background for each of these four 
physics elements contained in our model has been presented in Paper I. 
Simulation methodology for an expanding supershell model has also 
been described in Paper I. In this paper we present an ensemble of 
ideal three-dimensional, spherical expanding shells, in order to 
model the line emission of neutral ISM in massive star-forming 
regions in a starburst galaxy like M 82. Our instantaneous starburst 
model does not address issues related to more complicated geometry in 
order to understand how these shells are distributed in a galaxy, 
how they interact, or how the gas becomes available for fueling 
the massive star formation in our model GMCs. 

In this paper, we divide the evolutionary scheme of the expanding 
shell/GMC ensemble into two phases referred to as {\it Winds} and 
{\it post-SN}. The {\it Winds} phase begins with the formation of 
a star cluster and an H II region inside the GMC followed by the 
formation of a rapidly expanding hot bubble produced by strong 
stellar winds. This phase ends when the bubble breaks out of its 
parent GMC. In this phase, the parent GMC is assumed to be stationary, 
and acts as a dense uniform ambient ISM to the expanding shell formed by 
gas swept up by the bubble. The {\it post-SN} phase starts when the most 
massive star in the ensemble reaches its main-sequence lifetime, and 
explodes into a supernova. The shell expansion in this phase is first driven 
by repeated supernova explosions, then changes from pressure-driven 
({\it adiabatic}) to zero pressure ({\it non-adiabatic}) as the hot bubble 
cools. In the {\it post-SN} phase, the shell expands into a less dense 
uniform ambient ISM. A steady-state mechanical power and energy for each 
phase is assumed, in order to satisfy the requirements of the similarity 
model. Our non-LTE line radiative transfer method simply sums the 
line emission from the model shells and parent GMCs in the ensemble 
for the {\it Winds} phase, and just the model shells for the 
{\it post-SN} phase. Hence, our model is an idealistic approximation 
for a starburst galaxy; it may be considered the first step toward 
simulating the response of the gas environment in an evolving starburst 
region for the purpose of examining the effects of this evolution. 

\begin{figure}
\epsscale{1.0}
\plotone{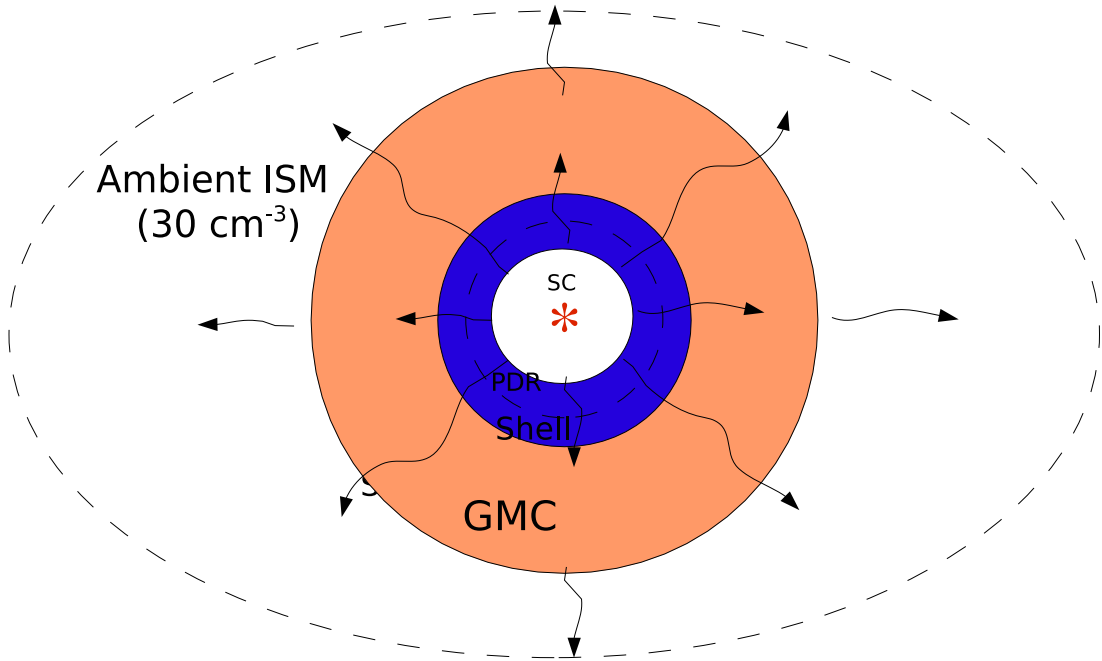}
\caption{This diagram illustrates the structural components associated
with a single star cluster within our model. The white region is 
the hot cluster wind, the blue region is the shell of material swept up 
from the giant molecular cloud, represented by the orange region.
The region exterior to the GMC is the ambient interstellar medium (ISM) 
with a fixed H$_2$ density of 30 cm$^{-3}$.
\label{isum}}
\end{figure}

A schematic diagram of the model components (PDR or shell and its parent 
GMC) is shown in Fig.~\ref{isum}. In this paper, we treat the PDR and 
shell as one gas component. The line intensity/flux for the shell and 
GMC components are calculated using the same method. For the {\it Winds} 
phase, the integrated line intensity/flux at each time step is the sum 
of line emission in the shell and its parent cloud. For the {\it post-SN} 
phase, the integrated line intensity/flux is the emission from the shell 
only. Due to the incomplete knowledge of the structure and physical 
state of the ambient ISM in a starburst galaxy, we do not include the 
molecular or atomic line emission from this component in our model. 
The justification of the ambient ISM density (30 cm$^{-3}$) will be 
discussed in \S~\ref{snparam}. In \S~\ref{m82}, we discuss the possible 
impact of this exclusion on the conclusions, specifically regarding 
the applications to M 82.

\subsection{Model Outline} \label{moutline}

Our basic model comprises a series of non-overlapping (i.e. non-interacting) 
spherical shells expanding into a uniform gas medium. These shells are 
all driven by winds from star clusters formed during an instantaneous 
starburst. The interior hot bubble is produced by stellar wind from 
an underlying super star cluster, whose properties are selected as 
discussed later. The thrust of our model simulation is to compute 
the molecular line emission from the swept-up shells and the associated 
parent GMCs. Since the underlying stellar radiation from the clusters 
has a pronounced effect on the properties of PDR regions of the shells, 
and since these properties are therefore also affected by the radius 
of the shell and evolutionary stage of the cluster, the SED of the 
molecular line emission from these shells contains a {\it signature} 
of the stage of evolution of the starburst. This variation with time, 
predicted by our model, offers a way of dating the starburst, at least 
in principle.

The set of our starburst models is divided into two phases, namely 
the {\it Winds} and {\it post-SN} phases. In the {\it Winds} phase, 
the shells propagate into their parent clouds, which are substantially 
more dense than the surrounding ISM in which they are embedded. 
In the {\it post-SN} phase, the shell breaks out of the parent 
cloud and expands into the uniform lower density ISM which pervades 
the entire galaxy. The same bubble/shell dynamical theory is used 
to describe the shell behavior in both phases. Since the simple 
similarity relations do not apply to a nonuniform ambient medium, 
we do not follow the shell expansion across the transition from 
cloud to the surrounding ISM. Instead the two phases are treated 
independently, and can be viewed as simply alternative 
scenarios. Continuity at the transition is only maintained in 
the mechanical luminosity of the wind ($L_{w}$) and the stellar 
luminosity ($L_{\ast}$) evolution of the central star clusters.
There is accordingly a {\it discontinuity} in the radius ($R_s$), 
velocity ($V_s$), and consequently the temperature ($T_s$), 
density ($n_s$), thickness ($d_s$), and the mass of each expanding 
shell across the boundary between the two phases. These quantities 
asymptotically approach those of a continuous model when the mass 
of the ISM swept up in the {\it post-SN} phase becomes greater 
than the mass of the parent GMC in the {\it Winds} phase. 
The {\it Winds} phase thus comprises younger and denser shells 
than in the {\it post-SN} phase. In principle, it is possible to 
model data with starbursts occupying a large range of potential 
ages and molecular gas excitation conditions, especially if the 
dense molecular core component were also included our model. 
For this study, our {\it Winds} phase model results are used 
only for a comparison with the observed expanding supershell 
surrounding SNR 41.9 + 58 in the center of M 82, which will be 
discussed later in \S~\ref{m82}. 

In each phase, the shell structure is computed with time as 
the independent variable. The final output dependent variables 
are the line fluxes (and profiles) for several molecules and
atoms each at a number of observed transitions, computed by
a non-LTE line radiative transfer code applied to each shell
and its parent cloud. The integrated line flux for each shell 
(and GMC) is the sum over the emission from the entire emitting
region. The total line flux for the shell ensemble is then the 
sum of the integrated line fluxes of all shells (and GMCs). 
Intermediate variables which determine these fluxes include the 
radius and velocity of each shell, its chemical structure, shell 
temperature and density structure, which are computed using a 
{\it Shell Dynamics} code, a stellar population synthesis code,
and a time-dependent PDR code. These codes are described in detail 
in the {\it Ph.D. Thesis} of Yao (2009, hereafter Yao Thesis, 
and references therein).

Our two-phase starburst model described above must also be 
characterized by a number of fixed parameters with adopted plausible 
values. These include, for example, the initial giant molecular cloud 
parameters (mass $M_{GMC}$, initial H$_2$ density $n_0$, and 
core H$_2$ density $n_c$), the star formation efficiency ($\eta$),
the star cluster related parameters (IMF, individual star mass $m_\ast$), 
the initial chemical composition of the parent clouds, and the density 
of the ambient ISM. These parameters, along with others, and 
their numerical values, are discussed in detail in the subsequent
sections, in association with the discussion of the PDR and radiative 
transfer codes. A brief summary of all variables and parameters are 
presented in Table~\ref{tbl3} at the end of this section. 

Finally, a chi-square ($\chi^2$) method will be used for fitting the 
model line spectral energy distribution to a set of molecular line 
data in order to estimate the starburst age(s) and observed total H$_2$ 
mass in the nuclear disk of M 82. In a subsequent paper (Yao 2009), we will 
present applications to luminous infrared galaxies (LIRGs) \citep[e.g.][]{yao03} 
beyond M 82 using our FIR/sub-mm/mm starburst model. We discuss the 
relationships between the excitation of CO molecule and SF properties 
of LIRG galaxies, and derive the behavior of the model CO-to-H2 conversion 
factor $X$ in a starburst galaxy. 

\subsection{Model Input Parameters} \label{initparam}

\subsubsection{Winds Phase} \label{windparam}

In 2005 Keto et al. observed $^{12}$CO(2-1) emission in the center of 
M 82 with a linear resolution of 17 pc at the source. They resolved 
$\sim$ 300 molecular clouds with masses ranging from $\sim$ 2 $\times$ 10$^3$ 
to 2 $\times$ 10$^6$ M$_{\odot}$. The mass spectrum of these GMCs
scales as $dN$/$dM_{GMC}$ $\propto$ $M^{-1.5 \pm 0.1}_{GMC}$, similar
to the Galactic one \citep{san85, sol87}. Keto et al. also found that
the mass spectrum of star clusters in M 82 follow the same power-law
distribution, suggesting that individual molecular clouds are 
transformed in the starburst into individual star clusters in their 
dense cores. Combining this result with other studies \citep[e.g.][]{wei05},  
we assume the GMC mass distribution responsible for the stellar 
outburst in our model has a power-law index of 1.5, and the mass ranges 
between 3.16 $\times$ 10$^3$ and 10$^7$ M$_{\odot}$. About 70\% of 
molecular gas mass in a model starburst will then be contained in the 
clouds with masses $>$ 10$^{6}$ M$_{\odot}$. It is also expected that 
much of the FIR luminosity due to star formation would arise from 
these massive clouds. To reduce the computation time, a discrete 
and arbitrary number of giant molecular clouds distributed similar 
to that discussed in Keto et al. (2005) is assumed. The masses 
for these discrete GMCs are 3.16 $\times$ 10$^3$, 10$^4$, 
3.16 $\times$ 10$^4$, 10$^5$, 3.16 $\times$ 10$^5$, 10$^6$, 
3.16 $\times$ 10$^6$, and 10$^7$ M$_{\odot}$, and are hereafter 
denoted as 3M3, M4, 3M4, M5, 3M5, M6, 3M6, and M7. The total number 
of clouds is about $\sim$ 400, and the total H$_2$ mass contained 
in the GMCs and shells is $\sim$ 1.69 $\times$ 10$^7$ M$_{\odot}$ 
with a total star clusters mass of 4.2 $\times$ 10$^6$ M$_{\odot}$. 
This selection is intended to provide a template cloud/cluster mass 
for scaling the model to fit the data for M 82. The best fitting 
molecular H$_2$ gas mass and the initial star cluster mass will be 
determined from a $\chi^2$ fitting method, as described later in \S~\ref{m82}. 

The average gas densities of GMCs in our Galaxy and starburst galaxies
are in the range a few 10 to a few times 10$^2$ cm$^{-3}$ 
\citep{dam86, jas92, wil08}, but their cores, where most of the stars 
form, have much higher densities. Higher gas densities are expected in 
more actively star-forming galaxies in accordance with the Schmidt law 
\citep{ken98}. Therefore, we adopt a value of 300 cm$^{-3}$ for the 
uniform initial H$_2$ gas density (i.e. $n_0$) for the M7 cloud based
on the densities for the most massive clouds in the study of molecular 
cloud properties in the active spiral M 51 by Scoville \& Wilson (2004). 
Since this work is proposed to be a complementary study to the dusty
starburst models developed by Efstathiou et al. (2000), we adopt the 
same core density namely $n_c$ = 2 $\times$ 10$^3$ cm$^{-3}$ for the 
calculations of the number of ionizing photons (used in computing 
the Str\"{o}mgren radius $R_S$ value). The radius of this 
10$^7$ M$_{\odot}$ cloud is 47 pc derived from the mass of the 
cloud and the assumed density with the assumed effective molecular 
weight $\mu$ = 2.36 \citep[e.g.][]{elm79, mak87}. 

We know that star formation takes place primarily in the dense cores 
of GMCs, but the details of the physical processes involved are not 
yet well understood. The efficiency of star formation (or the gas 
consumption rate) ranges from about 1\% in late-type spirals to 60\% 
or more in active star-forming galaxies \citep{ken98}. In this paper 
we adopt a moderate star formation efficiency $\eta$ = 25\% for our 
model starburst galaxies. A Salpeter IMF is assumed, i.e. 
$dN/dm_{\ast}$ $\propto$ $m^{-2.35}_{\ast}$ \citep[IMF;][]{sal55}, 
and the stellar mass is in the range 0.1 - 120 M$_\odot$.

We assume that the relationship between cloud mass and radius is 
the same as that derived from a CO survey for 273 giant molecular 
clouds in the Galactic inner disk by Solomon et al. (1987). From the 
measured relationship between the cloud size and the velocity line 
width, and the application of the virial theorem, they derived a 
power-law cloud density and mass relation, in which the mean gas 
density of the cloud is inversely proportional to the cloud size. 
Hence, the cloud mass is proportional to the square of the cloud 
radius (i.e. the mass surface density is a constant). From the 
studies of independent methods of determining the H$_2$ mass, 
Solomon et al. (1987) also demonstrated that these giant molecular 
clouds are bound principally by self-gravity and not by external 
pressure exerted by a hot phase of the ISM. Since we assume the 
mass distribution of GMCs in a starburst galaxy is similar to 
that in our Galaxy, we adopted the power-law relations of mass, 
radius, and density for our model GMCs as those defined in 
Solomon et al. (1987). This mass - radius relation has also been studied 
for GMCs in other galaxies, for example M 51 by Bastian et al. (2005), 
and has been found to be similar in form. Hence, the density and 
the radius for a model GMC having mass less than 10$^7$ M$_\odot$ 
can be written in the following forms,

\begin{eqnarray}
n_0 & = & 300 \ cm^{-3} \times \Big(\frac{R_{GMC}}{47 \ pc}\Big)^{-1}, \\
R_{GMC} & = & 47 \ pc \times \Big(\frac{M_{GMC}}{10^7 \ M_\odot}\Big)^{\frac{1}{2}}
\end{eqnarray}

where $R_{GMC}$ is the radius of the GMC with mass $M_{GMC}$ that is 
less than 10$^7$ M$_{\odot}$. Table~\ref{tbl1} summarizes 
the number distribution of the GMCs and their initial physical 
properties. For GMC mass less than 3.16 $\times$ 10$^3$ M$_{\odot}$, 
the predicted number of very massive stars ($>$ 30 M$_{\odot}$) in the 
star cluster is below 1.0. In addition the supernova wind will not be 
steady as assumed in our model, because of the relatively small numbers 
of contributing stars. Initially, the total GMC and cluster masses
are assumed to be $\sim$ 1.86 $\times$ 10$^7$ M$_{\odot}$ and 
$\sim$ 4.27 $\times$ 10$^6$ M$_{\odot}$, respectively.     

\begin{deluxetable}{lcccccccccc}
\tabletypesize{\scriptsize}
\rotate
\tablecaption{Initial conditions of GMCs and SCs in a modeling starburst system. \label{tbl1}}
\tablewidth{0pt}
\tablehead{
\colhead{GMC ($N_{norm}$)\tablenotemark{a}} & \colhead{log$_{10}$M$_{GMC}$\tablenotemark{b}} & \colhead{$n_0$\tablenotemark{c}} & \colhead{$n_c$\tablenotemark{d}} & \colhead{$R_{GMC}$\tablenotemark{e}} & \colhead{log$_{10}$$M_{SC}$\tablenotemark{f}} & & \colhead{$N_{\ast}$\tablenotemark{g}} & & \colhead{log$_{10}$$L^{\ast}_{SC}$\tablenotemark{h}} & \colhead{log$_{10}$$L^{mech}_{SC}$\tablenotemark{i}} \\
 & \colhead{(M$_\odot$)} & \colhead{(cm$^{-3}$)} & \colhead{(cm$^{-3}$)} & \colhead{(pc)} & \colhead{(M$_\odot$)} & \colhead{(0.1M$_\odot$ $\le$ $M_{\ast}$ $\le$ 120M$_\odot$)} & \colhead{($M_\ast$ $\ge$ 8M$_\odot$)} & \colhead{($M_\ast$ $\ge$ 30M$_\odot$)} & \colhead{(erg s$^{-1}$)} & \colhead{(erg s$^{-1}$)} 
}
\startdata
M7 (1.0) & 7.0 & 300 & 2000 & 46.8 & 6.4 & 7.1E6 & 2.2E4 & 5.0E3 & 42.8 & 40.1 \\
3M6 (1.77) & 6.5 & 534 & 3558 & 26.3 & 5.9 & 2.2E6 & 7.0E3 & 1.6E3 & 42.3 & 39.6 \\
M6 (3.1) & 6.0 & 949 & 6325 & 14.8 & 5.3 & 7.1E5 & 2.2E3 & 5.0E2 & 41.8 & 39.1 \\
3M5 (5.6) & 5.5 & 1688 & 11251 & 8.3 & 4.9 & 2.2E5 & 7.0E2 & 1.6E2 & 41.3 & 38.6 \\
M5 (10.0) & 5.0 & 3000 & 20000 & 4.7 & 4.4 & 7.1E4 & 2.2E2 & 50.0 & 40.8 & 38.1 \\
3M4 (17.7) & 4.5 & 5337 & 35578 & 2.6 & 3.9 & 2.2E4 & 70.0 & 15.0 & 40.3 & 37.6 \\
M4 (31.6) & 4.0 & 9487 & 63246 & 1.5 & 3.4 & 7.1E3 & 22.0 & 5.0 & 39.9 & 37.2 \\
3M3 (56.2) & 3.5 & 16876 & 112509 & 0.8 & 2.9 & 2.2E3 & 7.0 & 1.0 & 39.4 & 36.7 \\ \hline
Ensemble (127) & 7.27 & \nodata & \nodata & \nodata & 6.63 & 1.57E7 & 4.9E4 &1.1E4 & 43.1 & \nodata 
\enddata
\tablecomments{All data listed in the above table are for single SC, except for the Ensemble.}
\tablenotetext{a}{GMC type (number of GMC in an ensemble).}
\tablenotetext{b}{GMC mass.}
\tablenotetext{c}{Average gas density of a GMC.}
\tablenotetext{d}{GMC core density.}
\tablenotetext{e}{GMC radius.}
\tablenotetext{f}{Star cluster mass.}
\tablenotetext{g}{Star number for different stellar mass ranges.}
\tablenotetext{h}{Stellar bolometric luminosity.}
\tablenotetext{i}{Mechanical luminosity (Stellar wind + SN).}
\end{deluxetable}

\subsubsection{Post-SN Phase} \label{snparam}

All shells are propelled into a less dense ambient ISM during the {\it post-SN} 
phase. The intercloud medium of the central 1 kpc region of the Galaxy has been 
studied by Jog \& Solomon (1992), who find it to be mostly molecular with 
density between 30 and 100 cm$^{-3}$. Bally et al. (1988) also obtained 
an average molecular gas density of 50 cm$^{-3}$ for the region within 
a radius of 500 pc of the center of our Galaxy. In this paper, we assume 
a uniform ambient ISM with similar density surrounding the GMCs for 
our model.  

To investigate whether this medium should be considered as atomic 
or molecular, and to get an estimate of its mean density, we compare 
the observational constraints for various ambient ISM constituents 
(H$_2$, HI, and H II) for the central 1 kpc region of M 82. Table~\ref{tbl2} 
shows that gas with a column density of about $10^{23}$ cm$^{-2}$ is 
required by observations, and that the dominant state of the ISM is 
molecular. Hence, from this observed H$_2$ column density, and an adopted 
diameter of 1 kpc for the starburst region, we can derive the number 
H$_2$ density of about 30 cm$^{-3}$ (or a total H$_2$ mass of  
$\sim$ 1.2 $\times$ 10$^9$ M$_{\odot}$), and we adopt this figure for 
modeling the central 1 kpc region in M 82. 

In reality, intercloud gas in M 82 is unlikely to be uniformly 
distributed, as assumed in our model. Recent studies 
\citep[e.g.][and references therein]{gam07} show that a smoothly 
distributed turbulent medium consisting of atomic gas would quickly 
(within a few 10$^6$ yr) develop density fluctuations, becoming a 
highly non-uniform medium of molecular H$_2$, with density enhancements 
up to a factor of 100 or more times the mean density. For simplicity, 
we ignore these density fluctuations, and regard this medium as 
represented by its mean density, treating it as uniform for the 
purpose of computing the material swept-up by the expanding shells.  

\begin{deluxetable}{lll}
\tablecaption{Observed column density of various ISM components in the center of M82. \label{tbl2}}
\tablewidth{0pt}
\tablehead{
\colhead{Type} & \colhead{Column Density (cm$^{-2}$)} & \colhead{Reference} 
}
\startdata
H$_2$ & 6.1 $\times$ 10$^{22}$   & Wild et al. (1992) \\
      & a few $\times$ 10$^{23}$ & Mao et al. (2000) \\
HI    & 2.6 $\times$ 10$^{22}$   & Weliachew et al. (1984) \\
H II   & 9.0 $\times$ 10$^{22}$   & Carlstrom \& Kronberg (1991)
\enddata
\end{deluxetable}

The standard input parameters in our time-dependent PDR simulation that 
describe the physical properties of the model shell were summarized in 
Table 2 of Paper I. The cosmic-ray ionization rate $\zeta$ adopted in 
this paper (the standard Galactic value, i.e. 1.3 - 2.0 $\times$ 10$^{-17}$ 
s$^{-1}$) is up to two orders of magnitude lower than that measured in M 82 
\citep[e.g.][]{far94}. However, the main heating mechanisms of the PDRs 
are photoelectric, H$_2$ FUV pumping, and H$_2$ photodissociation. 
The contribution to the total heating rate from cosmic-ray ionization 
and excitation and the decay of turbulence within the cloud/shell is 
generally negligible. These processes only become important at large depths 
during late starburst evolution and/or dark cloud \citep{bel06}. 

The input parameters for a single-point dense dark-cloud model, which is 
used to produce the initial abundances at the first time step ($t$ = 0 yr) 
all depth steps for our PDR simulation, were also described in Paper I.
The dark-cloud assumption of chemistry is the same for all depth steps;
it is a reasonable guess for the initial gas conditions in a GMC
before star formation occurs. Our starburst model time step begins 
at $t$ = 1.0 $\times$ 10$^4$ yr, adopted as the time when the massive 
star formation occurs in the center of the GMC. For this time step and 
the subsequent time steps, the input abundances are re-set to the output 
abundances of the previous time step generated by the {\it UCL\_PDR} code. 
The chemistry at first iteration is calculated from gas temperature and 
attenuated FUV flux, and then revised iteratively until the balance 
criteria of heating and cooling is reached for each depth step at 
each time step. Hence, the final results are not significantly dependent 
on our initial dark cloud chemistry input at $t$ = 0 yr. The metallicity 
dependence appears in several key processes in the our PDR calculations, 
and accordingly we adopt solar abundances for the metals, i.e. unit 
metallicity. The dust-to-gas mass ratio is adopted as 1/100. 
Standard values of dust properties are used in the model (see Table 2.3 
in Bell 2006), though the {\it UCL\_PDR} code allows the various dust 
properties to be specified as free parameters which can vary with shell 
depth and time. 

The chemistry within the parent GMC outside the shell is also handled 
by the same PDR analysis, using the different (lower) density in this region. 
The incident FUV strength for the cloud region is the attenuated radiation 
field emerging from the outer boundary of the shell, and the FUV strength
inside the cloud is computed in the same way as for the shell, with the
computation of $A_V$ taking account of the lower density of the dust. 

Our multi-level non-LTE radiative transfer {\it SMMOL} code includes an 
empirical dust extinction model (see Table 1 in Mathis 1990). In our models, 
we assume the entire region containing all of the shells is unresolved. 
The lowest ten energy levels are incorporated for molecular species 
(CO, HCN, HCO$^+$, CN, HNC), three levels for atomic [C I] and [O I], 
and two levels for atomic [C II]. In Paper I, single collisional partner 
(H$_2$) was used in the radiative transfer calculations. In this paper,
multiple collisional partners (H, e$^-$, H$^+$, p-H$_2$, o-H$_2$, He) 
are taken into account in the statistical equilibrium equation 
calculation. The collisional excitation of molecular lines involves 
two partners, i.e. p-H$_2$ and o-H$_2$, but the excitation of [C I] 
fine structure lines is affected by collisions with all six 
particles, five (without He) for [O I] lines, and four (without 
He and H$^+$) for [C II] lines. Since these forbidden lines have 
very low radiative transition probabilities, the upper states
are populated primarily by collisions, and they are usually 
optically thin.

The input parameters to the {\it SMMOL} model are (1) molecular data 
including molecular mass, energy levels, transition frequencies, 
radiative rates and collisional rates; and (2) physical data describing 
the object to model. This includes the physical distance of the current 
grid point to the center of the shell, gas density, number densities of 
the six collisional partners (H, e$^-$, H$^+$, p-H$_2$, o-H$_2$, He), 
the fractional abundance of molecules or atoms, the gas (kinetic) and 
dust (thermal) temperatures, shell expansion velocity, and the 
microturbulent velocity. 

Table~\ref{tbl3} summarizes the parameters and variables used 
in our simulations.

\begin{deluxetable}{ll}
\tabletypesize{\scriptsize}
\tablecaption{Model parameters and variables \label{tbl3}}
\tablewidth{0pt}
\tablehead{
\colhead{Models} & \colhead{Description}   
}
\startdata
Independent Variable: & time or starburst age $t$ \\
Dependent Variables:  & shell radius $R_s$, expansion velocity $V_s$, number density $n_s$, and thickness $d_s$ \\
                      & gas (kinetic) temperature $T_{gas}$, dust (thermal) temperature $T_{dust}$ \\
                      & chemical abundances of different molecules and atoms in the shell \\  
                      & number densities of collisional partners H, e$^-$, H$+$, p-H$_2$, o-H$_2$, and He \\
Fixed Parameters:     & GMC mass $M_{GMC}$: 3.16 $\times$ 10$^3$  - 10$^7$ M$_{\odot}$ \\
                      & stellar mass $m_{\ast}$: 0.1 - 120 M$_{\odot}$  \\ 
                      & star formation efficiency (SFE) $\eta$ = 0.25 for {\it Winds}, $\eta$ = 1.0 for {\it post-SN} \\ 
                      & metallicity $\mathcal{Z}$ = 1.0 Z$_{\odot}$ \\ 
                      & gas-to-dust ratio = 100 \\ 
                      & ambient ISM density of each shell $n_{ism}$ (parent GMC at {\it Winds}, 30 cm$^{-3}$ at {\it post-SN}) \\
                      & microturbulent velocity $\delta$$v_D$ = 1.5 km s$^{-1}$ \\
Fitting Parameters:   & total gas mass of the ensemble $M_{total}$, burst age $t$ \\ \hline
Outputs:              & line profiles for each transition in each molecules and atoms \\
                      & integrated line intensity or flux  
\enddata
\end{deluxetable}

\section{Model Results} \label{ss}

In this section we present examples of our simulation results for 
the shell ensemble centrally illuminated by massive star clusters. 
Detailed results for individual expanding shells can be found in Yao et al. 
(2006) and Yao Thesis. These shells are modeled in a similar way for 
both {\it Winds} and {\it post-SN} phases as described in previous 
sections. A family of these evolving shells form the basis of our 
starburst models in accordance with our description in \S~\ref{esbm}. 
Applications of the shell ensemble to M 82 and more distant starburst 
galaxies will be presented in subsequent sections. 

The two modeling phases are indicated by {\it Winds} and {\it post-SN}
labels in tables and plots throughout the remainder of this paper. 

\subsection{Kinematics of The Swept-up Gas} \label{kint}

The strong stellar winds and supernova explosions from hundreds to 
thousands of the massive stars fuel the hot bubbles over a timescale 
$>$ 10 Myr. The kinetic energy in the supersonic wind is thermalized 
by a stand-off shock, and the high pressure downstream drives a strong 
shock into the ambient ISM. The swept-up gas condenses into a narrow 
shell as a result of radiative cooling. The wind mechanical luminosity 
$E_{mech}$ comes mainly from Wolf-Rayet (WR) stars, with some contribution 
from O stars. All other stars produce a negligible effect.

During the {\it Winds} phase, the sizes of the initial Str\"{o}mgren 
spheres in our model ensemble increase slowly with time. The Str\"{o}mgren 
radius ranges from 0.02 to 4.9 pc with the number of Lyman continuum photons 
between 1.5 $\times$ 10$^{49}$ and 5 $\times$ 10$^{52}$ s$^{-1}$ generated 
from the central star clusters derived from Equation (2) in Efstathiou 
et al. (2000). The wind bubble catches up with the ionization front 
of the compressed shell in a time less than 10$^5$ yr. The strong 
stellar winds cause the bubbles to expand quickly into their parent 
clouds and to sweep up more gas into the shells. When the most massive 
star in the most massive star cluster (i.e. 120 M$_{\odot}$ star in 
the M7 cloud) terminates as a supernova at $\sim$ 0.8 Myr \citep{mam88}, 
this marks the beginning of {\it post-SN} phase. At this time, the 
largest thin shell (M7) caused by the stellar winds is expanding at 
a speed of $\sim$ 50 km s$^{-1}$, and all the shells have swept up the
material in their parent clouds. The {\it Winds} phase ends earlier 
($<$ 0.8 Myr) for shells smaller than that for the M7 cloud. After 0.8 Myr, 
the shells begin to expand into a less dense uniform ambient ISM 
(i.e. 30 cm$^{-3}$). The mechanical energy produced by the first 
supernova and the subsequent ones re-energizes the shell formed 
in the {\it Winds} phase. 

The hot bubbles begin to cool at $\sim$ 0.8 Myr for the 3M3 shell 
and $\sim$ 7.5 Myr for the M7 shell. At this time, the radius and 
velocity of the M7 shell are about 270 pc and 24 km s$^{-1}$, 
respectively. After this time, the superbubbles start to lose 
their internal pressure, and the shell expansion velocity decreases 
rapidly. When the shell velocity approaches the sound speed of the 
ambient ISM, the shells should stall and become thicker and 
less dense. The latter effect is not included in our model, since 
the external pressure of the ISM is ignored. It is clear that 
the lifetime of the progenitor GMCs may be short, but the birth of 
massive star clusters and their impact on the surrounding ISM is 
profound. 

\begin{figure}
\epsscale{1.0}
\plotone{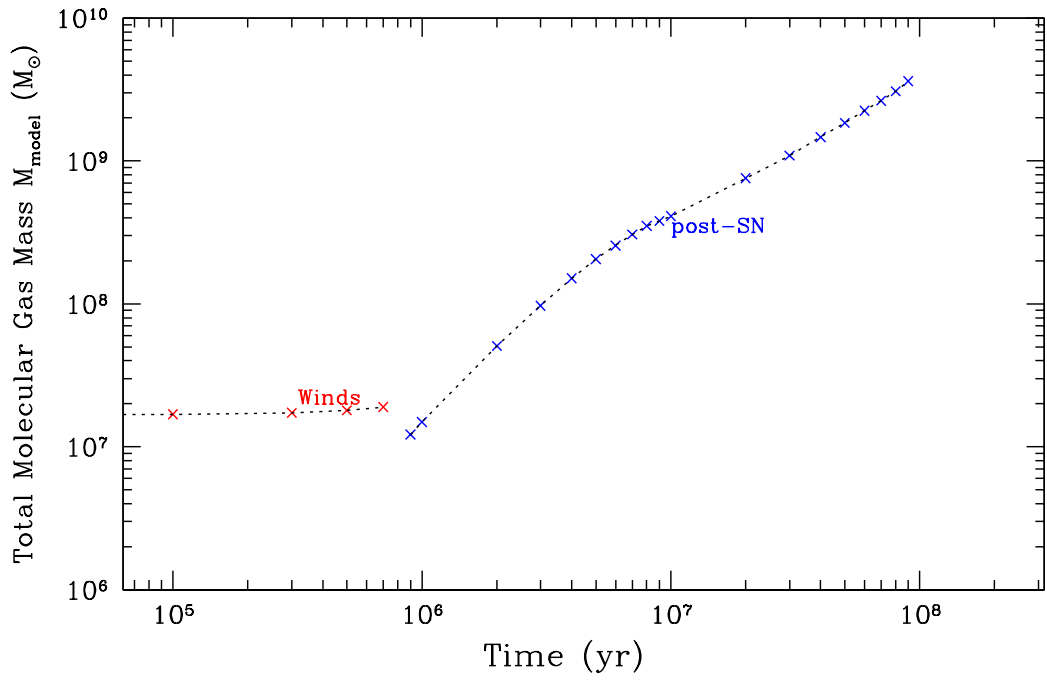}
\caption{Plot of the total H$_2$ mass swept-up by the shells and 
the mass remained in parent GMCs ({\it Winds} phase only) as a function 
of time. \label{totmh2}}
\end{figure}

In addition, we calculate the total amount of gas $M_{model}$ in
our model GMC/shell ensemble following the evolution of the shells, 
as shown in Fig.~\ref{totmh2}. The discontinuity seen at 1 Myr is 
caused by the phase change ({\it Winds} to {\it post-SN}), in which 
the parent GMC mass contained in the shell is no longer taken into 
account after the shell sweeps up all material in its parent GMC. 
This mass will be used as a template or reference value to be scaled 
to the total H$_2$ gas mass in a observed region of M 82 using 
a $\chi^2$ analysis for our model line SEDs, under the assumption 
that the line flux in the measured region is proportional
to the total molecular gas mass (see \S~\ref{m82} for details).

\subsection{Thermal Properties and Chemistry of the PDRs} \label{pdrs}

Over the 100 Myr of shell evolution the total mechanical wind power 
produced by individual cluster varies greatly, for example, from 10$^{34}$ 
- 10$^{40}$ erg s$^{-1}$ for the most massive cluster in our model ensemble. 
In this paper, the mechanical power profile is used only for obtaining 
average values over each phase in order to compute the shell dynamics. 
The kinetic energy of the shells is between 10$^{43}$ ergs and 10$^{54}$ ergs, 
depending on cluster mass. The FUV radiation incident on the inner 
surface of the shells ($A_V$ = 0) is a function of time ($G_0$ $\propto$ $R^{-2}_{s}$;
see in Paper I). The $G_0$ value is in units of the Habing field 
(1.6 $\times$ 10$^{-3}$ ergs cm$^{-2}$ s$^{-1}$) throughout this paper. 
This value decreases from about 10$^6$ - 10$^8$ (depending on cluster mass) 
at the onset of massive star formation (i.e. $t$ = 0 yr) to between 
10$^2$ and 10$^5$ respectively at 5 Myr when most of the massive stars 
($M_\ast$ $>$ 30 M$_\odot$) reach the end of their lifetime. At $t$ = 100 Myr, 
the $G_0$ values drop by 4 - 5 orders of magnitude. 

\subsubsection{Density and Temperature} \label{Gnd}

\begin{figure}
\epsscale{1.0}
\plotone{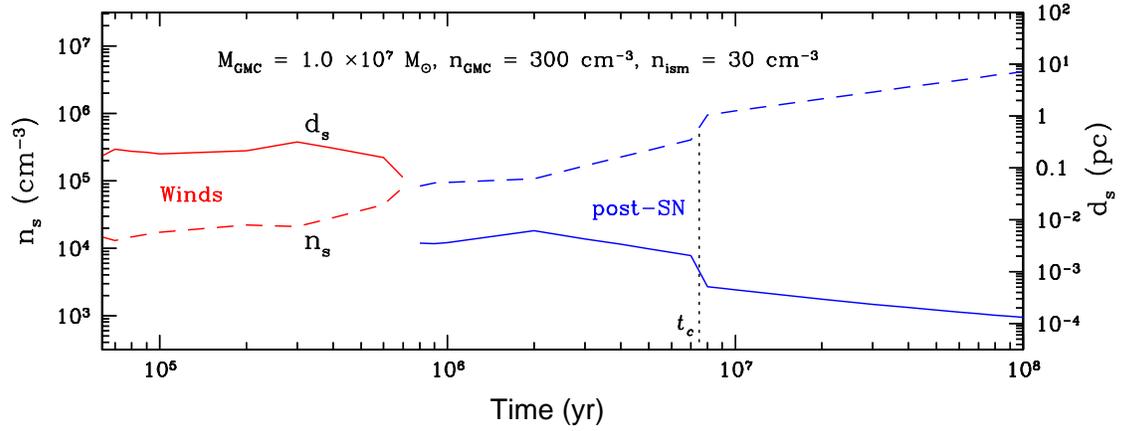}
\caption{Plot of the density ($n_s$, solid line) and thickness ($d_s$, 
dashed line) of an M7 shell as a function of time. The {\it Winds} phase is 
indicated by red curves, while the {\it post-SN} phase is indicated 
by blue curves. The radiative cooling of the hot interior occurs 
at $t_c$ indicated by the dotted lines.
\label{nd1}}
\end{figure}

The shell density, thickness, and temperature are calculated for both 
{\it Winds} and {\it post-SN} phases.  Fig.~\ref{nd1} shows an example 
of simulation results of an M7 shell. The density value varies from 
10$^2$ to 10$^6$ cm$^{-3}$ and the thickness is between 10$^{-3}$ and 
$\sim$ 10 pc over a 100 Myr period, depending on cluster mass. The 
{\it plateaus} seen at the beginning of the {\it Winds} phase are 
due to small changes in the expansion velocity and shell temperature. 
Before the shell sweeps up all of the material in its parent cloud 
($t$ $<$ 0.8 Myr), the shell density declines with increasing shell 
radius and decreasing shell velocity, and the shell thickness increases 
with time. The dense phase of the shells (10$^4$ - 10$^6$ cm$^{-3}$) 
is very short lived (between 10$^4$ - 10$^6$ yr). After the first 
supernova occurs (i.e. {\it post-SN} phase), the bubble continues 
expanding adiabatically into a lower density ambient ISM until a time 
$t_c$ (indicated in the plots), when this hot interior begins to cool 
and the shell enters the {\it snow-plow} phase while conserving its 
total momentum. The shell velocity then decreases rapidly with a 
corresponding decrease in shell density, to about three orders of 
magnitude lower than that at the {\it adiabatic} phase. Such large 
variation in the shell density is due to the range of dynamic pressure 
produced by the range in the shell expansion speed. The shell thickness 
increases from 0.1 pc at the beginning of the {\it post-SN} phase to 
10 pc at 100 Myr. Similarly, the thickness covers a large range because 
in the early phases, the shells are highly compressed and contain very 
little mass, so they are thin compared to later phases where these 
conditions are reversed.

In the plot, the first big jump occurs when the wind shock front 
catches up with the ionization front, and the expansion changes from 
H II to wind driven. For smaller GMCs, this transition takes 
place in less than 10$^3$ yr. The discontinuity (or gap) between 
{\it Winds} and {\it post-SN} phases is due to the model change 
from {\it Winds} to {\it post-SN} phase. A smaller jump is also seen 
when radiative cooling inside the bubble becomes dominant, the shell 
switches from the {\it adiabatic} to the zero pressure {\it snow-plow} 
phase. 

\begin{figure}
\epsscale{1.0}
\plotone{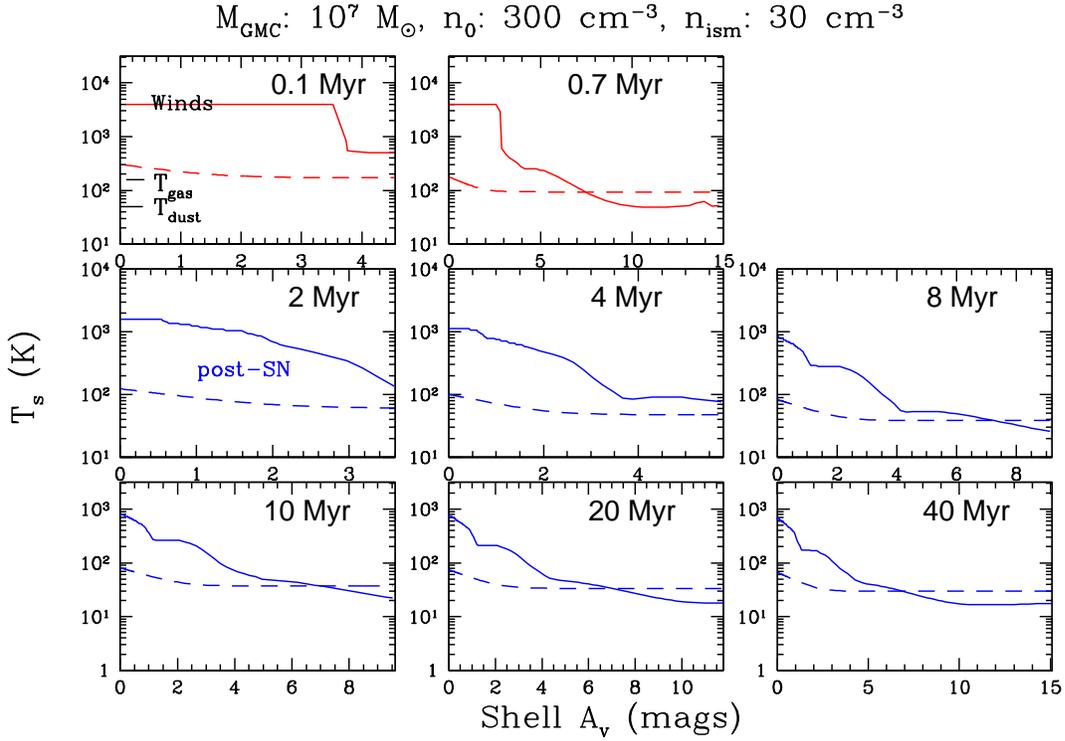}
\caption {Plot of the time-dependent gas and dust temperatures as a function 
of visual extinction $A_V$ for a M7 shell. Solid lines represent gas temperature, 
and dashed lines indicate dust temperature. The {\it Winds} phase model 
is indicated by red curves, and the {\it post-SN} phase model is indicated 
by blue curves.\label{T2Av1}}
\end{figure}

\begin{figure}
\epsscale{1.0}
\plotone{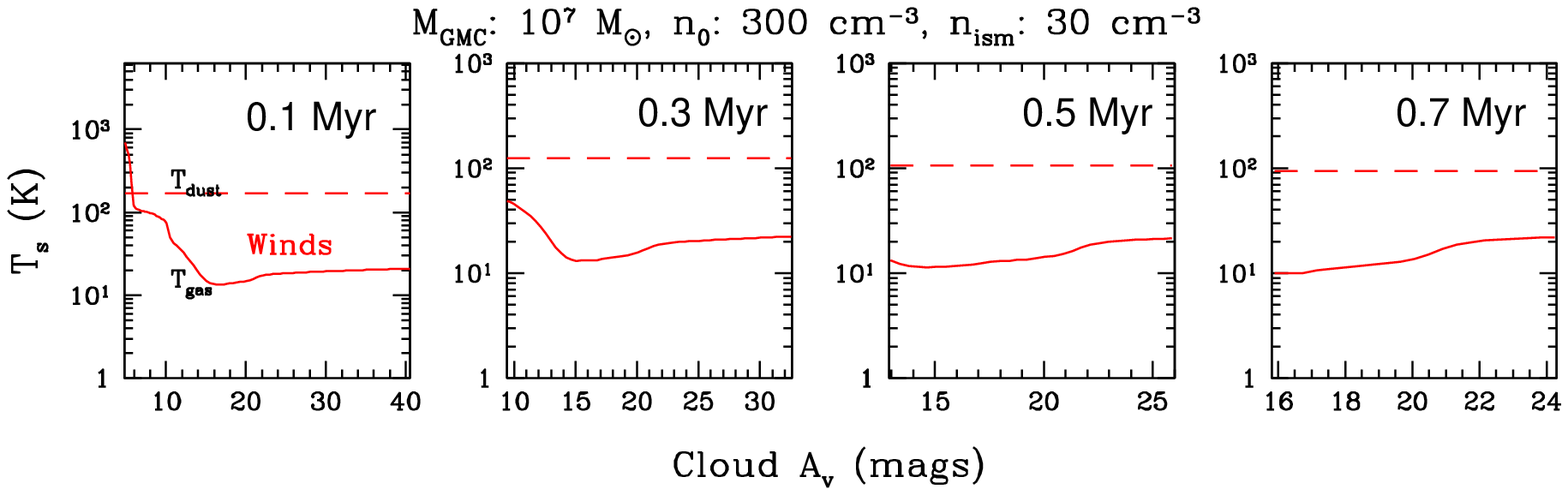}
\caption{Plot of the time-dependent gas and dust temperatures in the parent 
GMC (M7 cloud) as a function of visual extinction $A_V$. Solid lines 
represent gas temperature, and dashed lines indicate dust temperature.
\label{cT2Av1}}
\end{figure}

Fig.~\ref{T2Av1} shows an example of the gas and dust temperatures as 
a function of the visual extinction $A_V$ for an M7 shell at different 
starburst ages. The $A_V$ is set to be 0 at the inner surface of the 
shell (i.e. boundary between the hot bubble and the shell), and increases 
toward the outer edge of the shell (i.e. boundary between the shell and 
its parent cloud or the ambient ISM). During the {\it Winds} phase, 
the cloud $A_V$ progresses from the outer edge of the shell to the outer 
edge of the GMC (i.e. boundary between the GMC and its ambient ISM). 
The gas temperature has a negative gradient from the inner edges of the 
shells to the outer edges, because the FUV flux is attenuated owing to 
dust extinction resulting in decreasing photoelectric heating across 
the shells. The FUV field strength $G(t, A_V)$ at different $A_V$ 
(or depth in the shell) is a factor of $e^{-1.38 A_V}$ less than the 
flux at the surface of the PDR (or $G_0$). For example, at 1 Myr the 
FUV field strength at $A_V$ = 2 (layer of C$^+$/C/CO transition) is 
attenuated to $\sim$ 6\% of the value at the surface 
($G_0$ $\sim$ 10$^3$ - 10$^6$) for the shells in the ensemble. 
The gas temperature is in the range 10 - 1000 K across the shells. 
It is about 1 - 2 orders of magnitude higher than the dust temperature 
at the surface of the PDRs. Fig.~\ref{cT2Av1} shows an example of the 
temperature structure for an M7 cloud before the shell sweeps up all 
of its materials. The parent cloud is also heated by FUV radiation 
from the central star cluster. The minimum $A_V$ for the GMC corresponds 
to the extinction due to the shell at the shell-cloud interface, and 
the maximum $A_V$ is the extinction at the outer edge of the cloud. 
The gas temperature changes from 1000 K to about 10 K across the clouds, 
it drops rapidly with $A_V$ once the FUV field is attenuated, reaches 
a minimum when the CO cooling is most efficient. But when the gas 
becomes optically thick toward the outer edge of the parent GMC, 
the gas temperature rises slightly with increasing $A_V$ (e.g. age 
beyond 0.3 Myr) due to the assumption of a semi-infinite slab
used in our PDR model. Since the cooling line emission does not 
escape from the outer edge of the GMC, the cooling of the gas is 
inhibited. This causes an increase in the equilibrium gas temperature 
at large $A_V$. Although this effect is nonphysical, it affects only 
our {\it Winds} model, and it has no effect on the {\it post-SN} phase, 
and thereby no effect on our derived age discussed later in \S~\ref{m82}.

Although the physical properties of each giant molecular cloud and 
the star cluster born in its center vary greatly with cluster and 
cloud mass, the model profiles for the shell density, thickness 
and temperature are similar. This implies that different initial 
cloud conditions in a starburst environment may yield similar 
gas properties through the entire evolution. However, for individual 
shells, the physical properties of gas inside the shells change 
drastically with time. These gas properties that contain the imprint 
of different evolutionary phases, also determine the molecular line 
radiative transfer, and hence the spectral energy distribution of 
line fluxes. It allows us in principle to date the burst age by 
modeling the line spectrum energy distribution for various molecular 
tracers and comparing them with the observations of a starburst galaxy. 

\subsubsection{Chemical Evolution} \label{abund}

\begin{figure}
\epsscale{1.0}
\plotone{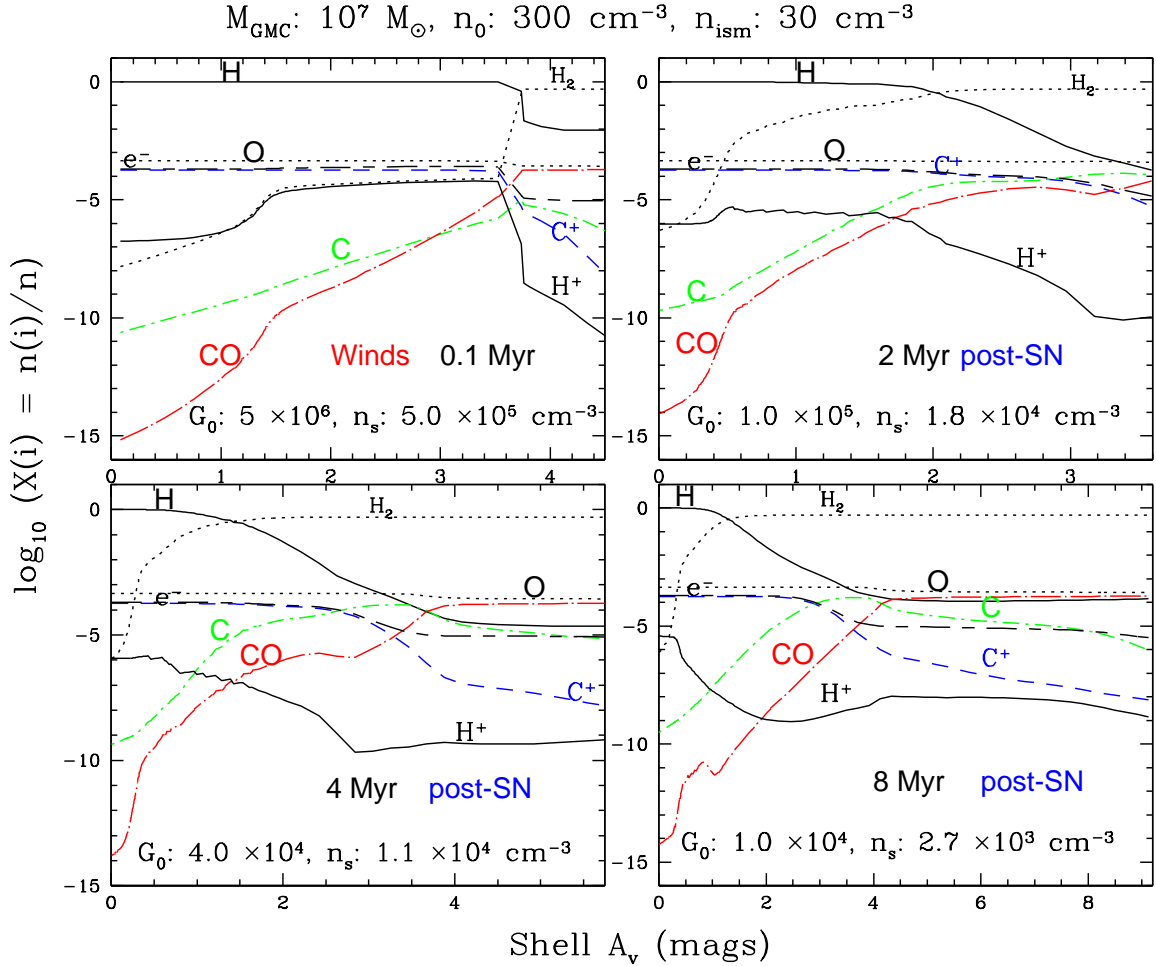}
\caption{Plots of the time-dependent chemical abundances of the main species 
(H, H$_2$, H$^+$, e$^-$, C, C$^+$, O, and CO) relative to the total hydrogen 
density, as a function of visual extinction $A_V$ for an M7 shell. 
\label{x2Av1}}
\end{figure}

\begin{figure}
\epsscale{1.0}
\plotone{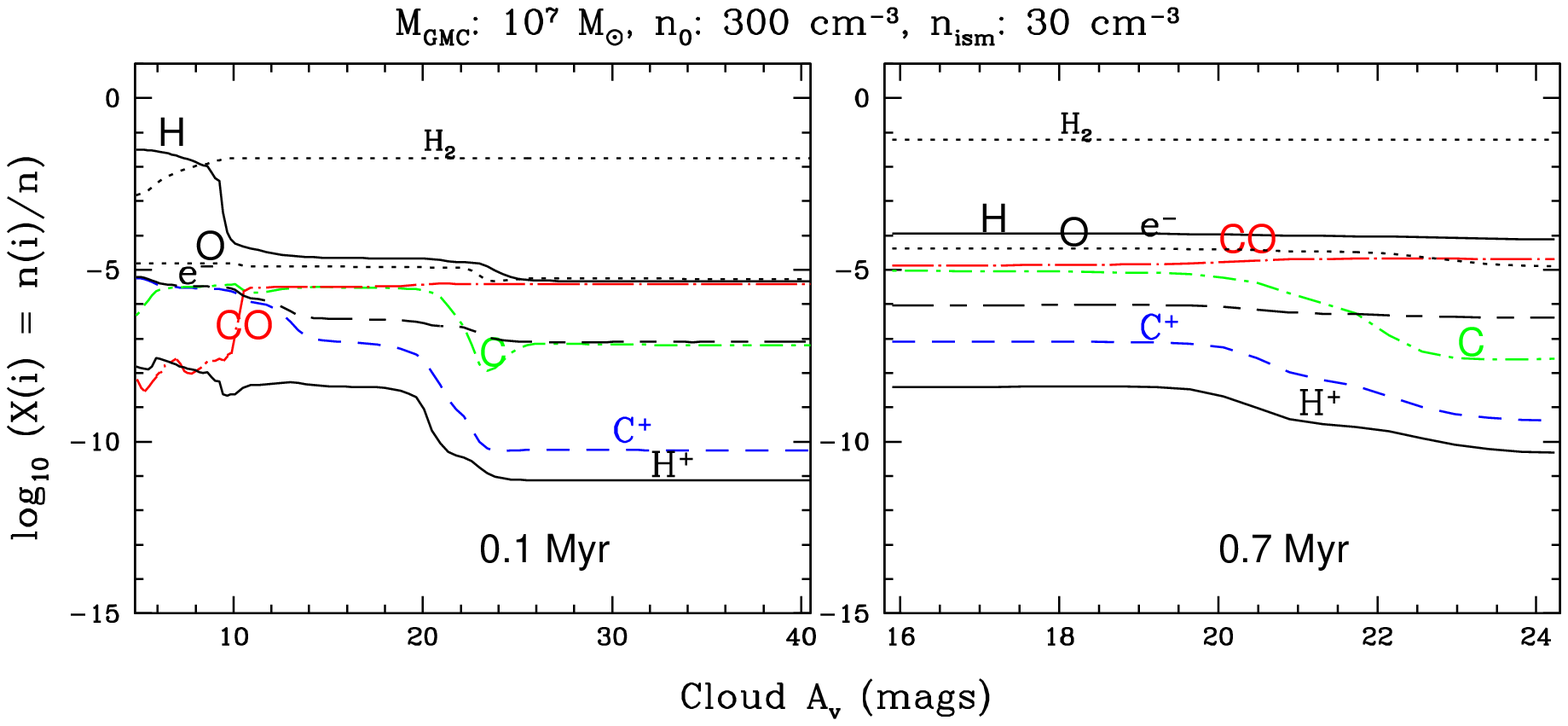}
\caption{Plot of the time-dependent chemical abundances of the main species 
(H, H$_2$, H$^+$, e$^-$, C, C$^+$, O, and CO) relative to the total hydrogen 
density, as a function of visual extinction $A_V$ for the most massive GMC M7 
in the ensemble.
\label{cx2Av1}}
\end{figure}

The chemical structure inside the shell is stratified. The FUV photons 
are gradually absorbed and lead to relatively sharp transitions. 
In Fig.~\ref{x2Av1}, the transitions of atomic species (H$^+$/H, C$^+$/C, 
O) to molecular gas (H$_2$ and CO) in an M7 shell are shown. The H$_2$ 
abundance becomes much more enhanced at $A_V$ $>$ 1, and the formation of
CO occurs at $A_V$ = 3 - 4. At the surfaces of the shells, the dominant 
coolant is the [O I] 63 $\mu$m fine-structure line. Deeper into the shells 
and the clouds the cooling by [C II] 158 $\mu$m, [C I] 610 $\mu$m, and 
CO becomes dominant (not shown). The chemical structure inside the shells 
changes significantly for the first few million years. This further 
justifies the use of a time-dependent PDR model for our shell evolutionary 
models. Figs.~\ref{cx2Av1} shows the chemical evolution inside an M7 cloud, 
before the shell sweeps up all of its material. 

\subsection{FIR/sub-mm/mm Line Emission in Individual Shells and A Shell Ensemble} \label{lrad_ss}

\subsubsection{$^{12}$CO and Its Isotope $^{13}$CO} \label{co}

\begin{figure}
\epsscale{1.0}
\plotone{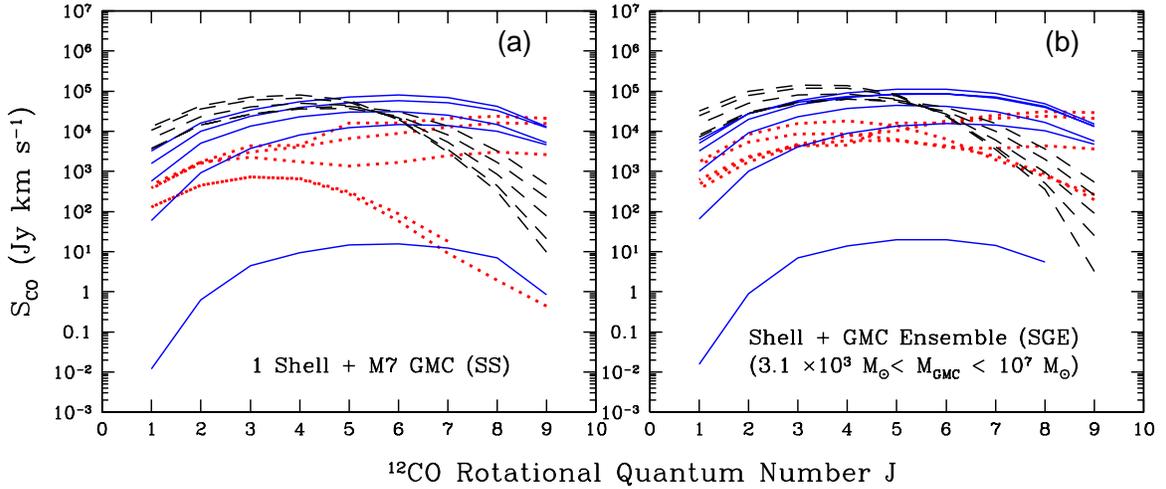}
\caption{Plots of model $^{12}$CO line SEDs for two different 
configurations. The {\it Winds} phase models are indicated by 
red dotted lines ($t$ $\le$ 0.7 Myr), while the {\it post-SN} 
phase models are indicated by blue solid lines (1 $\le$ $t$ $<$ 8 Myr) 
and black dashed lines (8 $\le$ $t$ $<$ 100 Myr).
\label{lsed_co1}}
\end{figure}

Fig.~\ref{lsed_co1} shows our model $^{12}$CO line SEDs ($J$ = 1$\ldots$9)
as a function of the starburst age for a single shell (corresponding to 
the M7 GMC) and a shell ensemble. In plot (a) the total line flux 
$S_{CO}$ is summed from gas in one single expanding shell and its parent 
GMC with a mass of 10$^7$ M$_\odot$ (M7 GMC and Shell or SS model). 
The parent GMCs contribute significantly to the total lower-$J$ line 
emission during the {\it Winds} phase. Our model results show during 
earlier {\it Winds} phase about 50\% - 100\% of total the $^{12}$CO(1-0) 
line emission comes from the M7 cloud, but it decreases to less than 
24\% at the $^{12}$CO(5-4) line, and to almost no contribution at $J$ $>$ 5. 
Our model M7 cloud has lower density than lower mass clouds (i.e. 3M3 - 3M6) 
and hence is less effective at exciting higher $J$ transitions. 
The nonphysical small rise in $T_{gas}$ at high $A_V$ (see Fig.~\ref{cT2Av1})
has almost no effect on the CO (1-0) line flux. At around 1 Myr, 
the line intensity drops by three orders of magnitude 
because we have not included the gas swept up in the GMCs in the 
subsequent model of the shells (i.e. in the {\it post-SN} phase). 
These model line fluxes predicted for individual expanding shells 
can be used as a comparison with future observations, for example, 
the known expanding supershell centered around SNR 41.9 + 58 in M 82, 
in order to constrain the physical conditions of the gas and the 
age of individual shells. 

In plot (b) of Fig.~\ref{lsed_co1} the total line flux $S_{CO}$ is 
calculated from all shells and their parent clouds in an ensemble with 
3.1 $\times$ 10$^3$ $\le$ $M_{GMC}$ $<$ 10$^7$ M$_\odot$ (Shell $+$ GMC 
Ensemble or SGE model). If multiple transitions CO data for individual 
expanding shells become available in the near future, models presented 
in plot (a) could be useful to constrain the burst age and gas mass in 
the shell, such as the supershell described in previous section.
More than 80\% of the $^{12}$CO line emission arises from the massive 
shells (3M5 - M7) in the ensemble. The line SEDs have two distinct maxima 
with one near the $J$ = 6 - 5 transition and another near the $J$ = 3 - 2 
transition. The first maximum is associated with burst age between 0.3 and 
7 Myr, and the second maximum is mainly associated with age older than 
7 Myr. At age 0.2 Myr, the two maxima (4 - 3, 8 - 7) seen in the line SEDs 
are due to the sum of line emission of gas in the shells and parent clouds. 
It is clear that the CO excitation in the line SEDs varies with shell 
expansion or starburst ages. At $\sim$ 1 ($\pm$ 0.2) Myr ({\it Winds} 
and {\it post-SN} phase transition), the $S_{CO}$ is a few orders 
of magnitude lower than those for other ages. This is an artifact of 
the switch from {\it Winds} to {\it post-SN} phase, where the GMC mass 
swept up in the {\it Winds} phase is not carried forward into the 
{\it post-SN} phase, and the continuity equation (or mass conservation) 
is applied to the less dense ISM (i.e. $n_{ism}$ = 30 cm$^{-3}$) 
instead of the GMC. Plot that shows the model line SEDs of $^{13}$CO,
HCN, HCO$^+$, and Table that calculates the fraction of $^{12}$CO,
HCN, and HCO$^+$ line emission from individual shells and their parent 
clouds are presented in Yao Thesis.

\subsubsection{Atomic C, O, and C$^+$} \label{atom}

The atomic forbidden transitions are the most important cooling lines 
arising in PDRs. The ratios of these lines and CO cooling lines can be 
used to derive the physical conditions in PDRs: for example, the incident 
FUV flux $G_0$, gas density $n$ and temperature $T_{gas}$, as well as 
the ratio of $G_0$/$n$. By comparing these model line ratios with 
observations, we can constrain the physical properties of atomic gas 
within a detected region. 

\begin{figure}
\epsscale{1.0}
\plotone{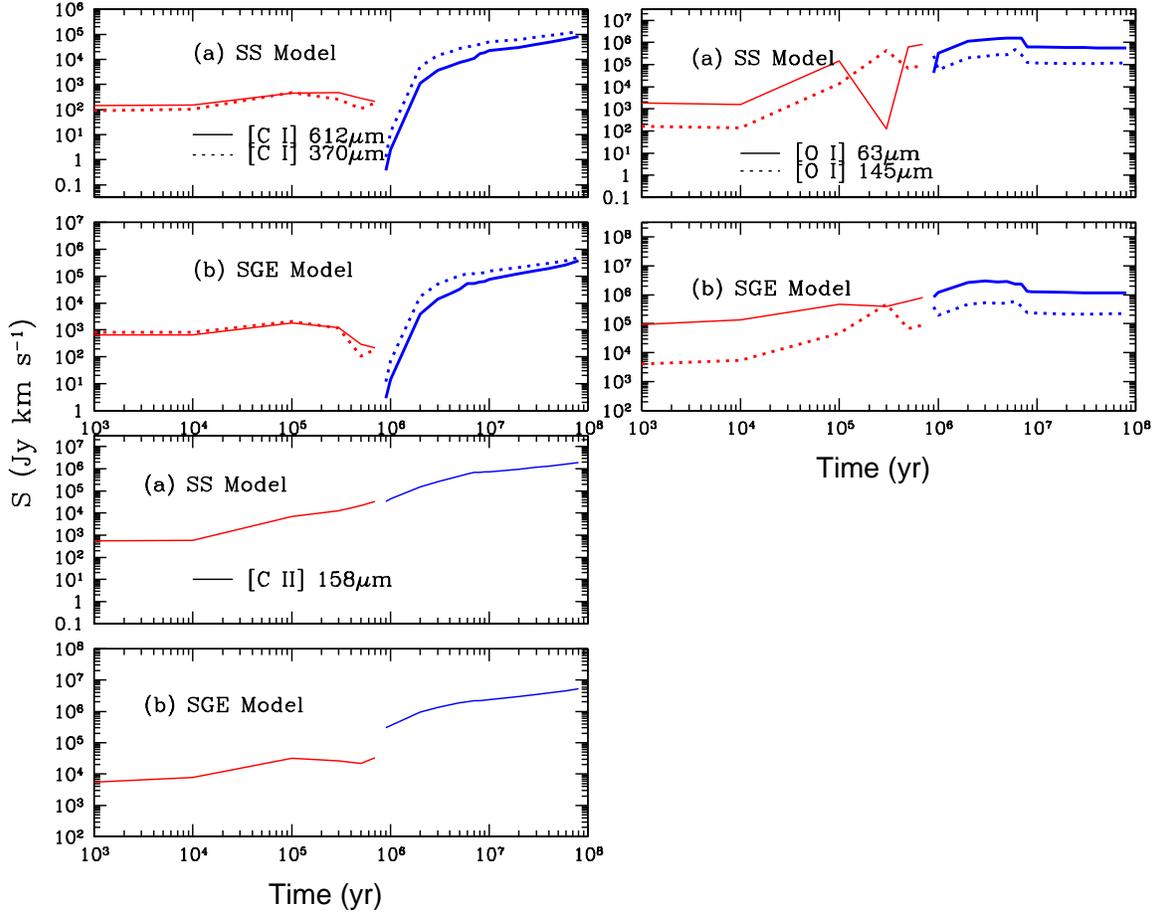}
\caption{Plots of model atomic line fluxes (C, O, C$^+$) as a function 
of time. Upper left plots (a) and (b): the solid curves are the [C I] 609 $\mu$m
lines, and the dashed curves are the [C I] 370 $\mu$m lines. Upper right plots (a)
and (b): the solid curves are the O I] 63 $\mu$m lines, and the dashed curves are 
the [O I] 145 $\mu$m lines. Lower left plots (a) and (b): model [C II] 158$\mu$m line 
flux as a function of time. The red color indicates {\it Winds} model, and the 
blue color indicated {\it post-SN} model. 
\label{lsed_c}}
\end{figure}

Upper left panels of Fig.~\ref{lsed_c} show the model line flux (in 
Jy km s$^{-1}$) for shell/GMC ensemble as a function of time for the 
[C I] 370$\mu$m line at 809.3 GHz and [C I] 609$\mu$m line at 492.1 GHz 
for the most massive shell/GMC and the shell ensemble. The 
[C I] 370$\mu$m transition has an excitation temperature of 63 K
and a critical density of 3 $\times$ 10$^3$ cm$^{-3}$. Both values
are higher than the [C I] 609$\mu$m transition (24 K and 
5 $\times$ 10$^2$ cm$^{-3}$). The atomic line fluxes along with the 
molecular line fluxes that we presented here are for the template model 
of the star clusters and molecular H$_2$ clouds. The values for the actual 
masses for M 82 will be derived from a fit of the fluxes of this template 
model to the observed fluxes. The [C I] 370$\mu$m line flux emitted in 
the shells is generally higher than that in the [C I] 609$\mu$m line, 
whereas in the cooler less dense parent clouds seen during the {\it Winds} 
phase the [C I] 370$\mu$m to [C I] 609$\mu$m ratio is close to one 
(as seen in both plots (a) and (b)). The massive shells are the dominant 
source for the neutral carbon line emission in the {\it post-SN} phase. 
The discontinuity seen in the plots (near 1 Myr) is a result of switching 
phase from {\it Winds} to {\it post-SN} as explained previously in 
connection with molecular emission. Both line fluxes increases with time 
in the {\it post-SN} phase. The integrated fluxes of atomic lines
predicted by our model generally increases with time, because more 
gas has been swept into the shells.

Upper right panels of Fig.~\ref{lsed_c} show the model line fluxes as a 
function of time for [O I] 63 $\mu$m line at 4744.8 GHz and [O I] 145 $\mu$m
line at 2060.1 GHz. The intensity increases with time for both lines, and 
then levels off after 8 Myr due to a sub-thermal excitation, the 
critical densities of these two lines are above 10$^5$ cm$^{-3}$. The 
[O I] 63 $\mu$m line flux is clearly stronger than the [O I] 145 $\mu$m 
line throughout the entire starburst evolution. The [O I] 63 $\mu$m 
transition has a lower excitation temperature (228 K) than the [O I] 145 $\mu$m 
transition (326 K), and hence it is easier to be excited.

Lower left panels of Fig.~\ref{lsed_c} show the model line flux of 
[C II] 158$\mu$m line at 1900 GHz as a function of time. The [C II] 158$\mu$m 
line has an excitation temperature of 92 K and a critical density of 
3 $\times$ 10$^3$ cm$^{-3}$. The flux of this line increases with time.

\section{Understanding of Molecular Gas and Starburst Ages in M 82} \label{m82}

In this section, we apply our evolving starburst models by comparisons
to an expanding molecular supershell centered around the supernova remnant
SNR 41.9 + 58 in the starburst galaxy M 82, and to the multiwavelength 
data of the central 1 kpc regions of M 82, in order to arrive at some 
conclusions about the nature of these two regions.

\subsection{The Supershell Surrounding SNR 41.9 + 58} \label{m82ss}

Observations have detected an expanding supershell centered around 
the bright SNR 41.9 + 58 in both molecular line and radio continuum 
\citep[e.g.][]{wei99, wil99}. This supershell has a diameter of 
$\sim$ 130 pc, an expansion velocity of $\sim$ 45 km s$^{-1}$, and a 
mass of $\sim$ 8 $\times$ 10$^6$ M$_{\odot}$. The kinetic energy of 
the observed supershell is estimated to be about 1.6 $\times$ 10$^{53}$ ergs 
\citep{wei99}. The kinematic evidence for the supershell appears most 
readily in the $^{13}$CO(1 - 0) position-velocity (PV) plot \citep{nei98} 
as a depression on the west side of M 82, bounded by a feature emerging 
toward lower velocities and possibly blended with emission associated 
with gas following orbits in the bar potential. Neininger et al. (1998) 
conclude that the depression seen in the $^{13}$CO(1-0) PV plot coincides 
with peaks in emission of [Ne II] and radio recombination lines, providing 
evidence that the void is populated by ionized gas inside the supershell. 
Seaquist et al. (2006) show that their PV plot reveals no depression in 
$^{12}$CO $J$ = 6 - 5 but instead find a region filled with $^{12}$CO 
$J$ = 6 - 5 emission that is not evident in the underlying $^{12}$CO 
$J$ = 1 - 0 map. Their line ratio PV map is consistent with the 
appearance of the channel maps, which show emission in the shell region 
extending over a very broad range in velocity. Seaquist et al. (2006) 
conclude that the location of this supershell contains CO with higher 
than average excitation, together with the ionized gas. The cavity 
created by the supershell is  not associated with prominent emission in 
higher density tracer such as HCN and HCO$^+$ in their low-excitation 
lines \citep{bas93, sea98}. This implies that the higher state of 
excitation may be due to higher kinetic temperature. Besides the known 
expanding supershell centered around SNR 41.9 + 58, there is evidence 
for other shells having sizes from several tens of parsecs to more than 
1 kiloparsec, and kinetic energies between $\sim$ 10$^{50}$ and 
10$^{55}$ ergs \citep[e.g.][]{lok87, gar01, wil02, bar05, bay08}.

\begin{figure}
\epsscale{1.0}
\plotone{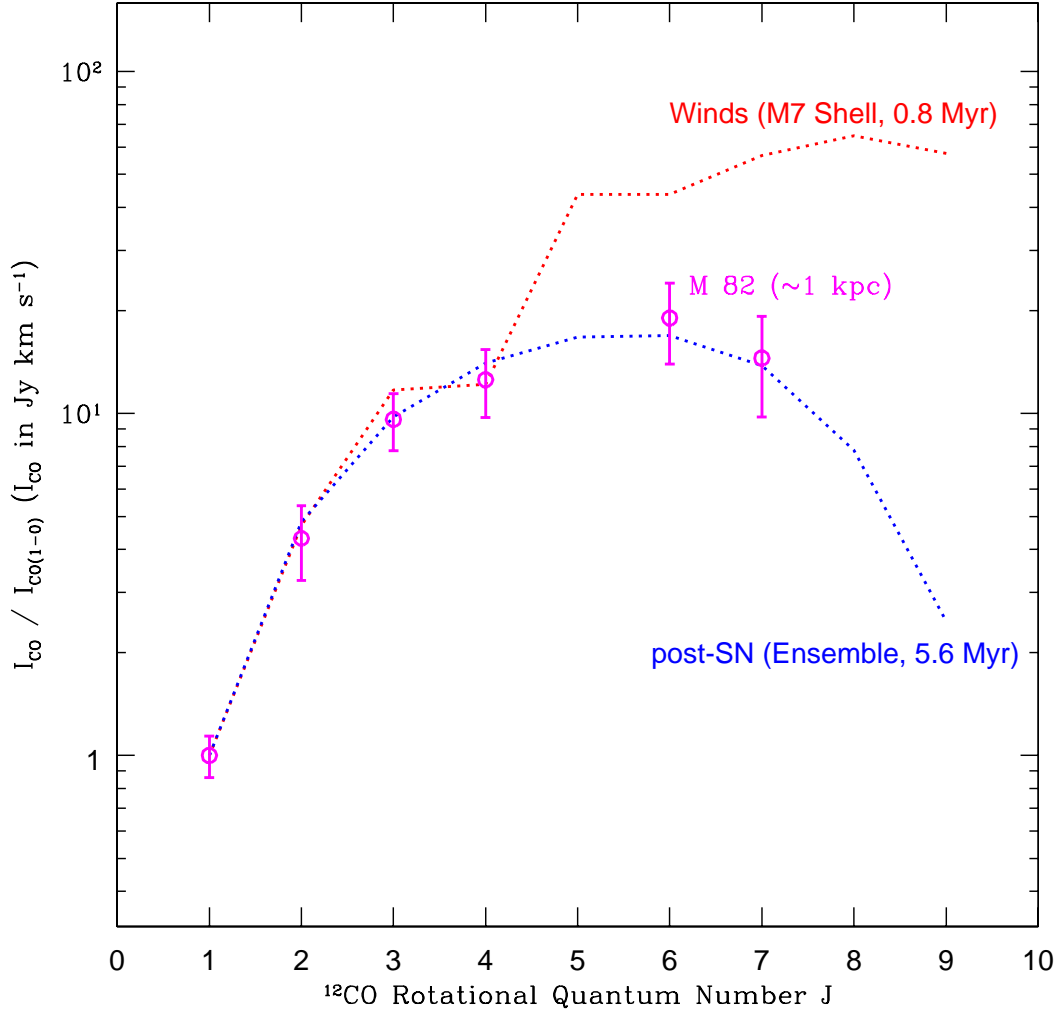}
\caption{Plot of model line ratios of $^{12}$CO high $J$ transitions to the (1-0) 
transition (i.e. $I_{CO}$/$I_{CO(1-0)}$, $I_{CO}$ in units of K km s$^{-1}$) as 
a function of rotational quantum number $J$ for an expanding supershell 
(M7) at age 0.7 Myr ({\it Winds} phase, red dotted curve). For comparison, 
a similar plot is shown of the observed (magenta circles with errorbars) 
and modeled SED (blue dotted curve) of the central $\sim$ 1 kpc region of 
the disk of M 82 to represent the background disk emission with lower 
excitation. For details of the latter model fit, see \S~\ref{lsed}.
\label{ssco}}
\end{figure}

The comparison of the kinetics of our single shell model with the 
observed supershell in M 82 is summarized in Table 3 of Paper I. 
Our model results and the observations agree remarkably well.
In this paper, we investigate the state of excitation of the molecular gas
in the supershell relative to that of the surrounding CO emitting gas
in M 82, by comparing the predicted line ratios in the shell to those 
in the surrounding gas. For the surrounding gas we use line ratios 
computed for the bulk of the disk molecular gas based on our forthcoming 
analysis of fitting our model for a shell ensemble to the observed 
line ratios for the central 1 kpc (see \S~\ref{lsed}). Fig.~\ref{ssco} 
shows this for the line ratios of $^{12}$CO high $J$ transitions to 
the (1-0) transition (i.e. $I_{CO}$/$I_{CO(1-0)}$, $I_{CO}$ in units 
of K km s$^{-1}$) for the model supershell. 
The jump in the ratios seen at $J$ = 3 to 5 results from the addition 
of line emission of M7 shell to that of its parent cloud, where the 
GMC contributes 5 - 45\% to the total line emission for $J$ $\le$ 4, 
but less than 0.2\% for $J$ $>$ 4. The plot shows clearly that our 
model for the supershell (red dashed curve) predicts that its line 
SED exhibits a higher level of excitation than the surrounding emission 
within M 82 (represented by the adjacent curve). Thus, one can expect 
some excess emission at high excitation transitions in the supershell 
after the underlying low excitation is subtracted out. Our model 
results are in qualitative agreement with the observational evidence 
for higher than average excitation emission in the supershell 
\citep[e.g.][]{nei98, sea06}. When higher quality and more extensive 
data on the excitation become available, our model predictions can be 
useful in interpreting the observations. 

\subsection{FIR/Sub-mm/mm Line Emission in The Central Region} \label{center}

Here we use the model components described in \S~\ref{ss} to produce
a fit of our model line spectral energy distribution to the 
observations of molecular gas in the central 1 kpc region. The purpose 
is to determine whether it is possible to model the FIR/sub-mm/mm 
line emission in a massive star-forming galaxy, and whether there 
is a relation between the molecular gas properties and the age of 
the starburst (i.e. finding the {\it age indicator}), and to assess the 
overall impact of the starburst on the fine scale structure and 
physical conditions of the ISM in M 82. 

We treat the entire central 1 kpc as an evolving starburst region, 
which can be modeled by following the evolution of an ensemble 
of expanding shells and clouds. Hence, different gas chemistry 
scenarios can be simultaneously at play in the center of this galaxy. 
However, our model does not attempt to reproduce or model the geometrical
distribution of shells in an actual starburst system. In any event 
this distribution is unknown since the individual shells are not 
observed. The total line emission is assumed to be represented by 
the sum of the emission from all the shells in the model ensemble, 
which will then be used to compare with the observed data to 
estimate the stellar mass, the total H$_2$ mass, and 
the age of the associated starburst in the measured region.               

\subsubsection{Observational Data} \label{m82obs}

The central concentration ($\sim$ 1 kpc) of molecular gas in M 82, 
which feeds the strong star formation activity, has been studied by 
many authors since the 1980s \citep[e.g.][]{yas84, wil92, gus93, wei99, 
mao00, paw00, wei01, war03}. Interesting results arise from these 
studies. For example, the observed CO line SED and line ratios can 
be reproduced by emission from low ($n$(H$_2$) $\approx$ 10$^{3}$ cm$^{-3}$) 
and high ($n$(H$_2$) $\approx$ 10$^{3.5-4.5}$ cm$^{-3}$, $T_{kin}$ $\ge$ 40 K) 
excitation gas components using a Large Velocity Gradient (LVG) method 
\citep[][and references therein]{wei05}. The high excitation component, 
responsible for the excitation of levels 
beyond $J$ = 4, arise from dense and warm gas, while the low excitation 
component is emitted by diffuse low density gas. The LVG method assumes 
a uniform abundance and velocity gradient across the modeling region, 
and no star formation history is considered as a cause for these conditions. 
It is this singular distinction which is the focus of this paper. 

The excitation conditions of multiple transitions of dense gas tracers 
HCN and HCO$^+$ in M 82 have also been investigated, for example, by 
Seaquist \& Frayer (2000). It was found from an LVG model, that both 
species are excited under a common set of conditions in star-forming 
regions where the $n$(H$_2$) is near 10$^5$ cm$^{-3}$, $T_{kin}$ = 50 K, 
and the abundances of HCN and HCO$^+$ are 2 $\times$ 10$^{-8}$ and 
1 $\times$ 10$^{-8}$, respectively \citep{saf00}. Molecular lines are 
commonly observed at 22$^{\arcsec}$ beam size, which covers about 680 pc 
of the center with a total H$_2$ mass of a few times 10$^8$ M$_{\odot}$
in M 82. 

The atomic coolant, far-infrared lines in M 82, e.g. [C I] 370 $\mu$m, 
609 $\mu$m, [O I] 63 $\mu$m, 146 $\mu$m, and [C II] 158 $\mu$m, have 
been studied by several groups 
\citep[e.g.][and references therein]{stu97, col99, paw01, lor96, neg01}.  
As is the case for the molecular lines, the ratios of these cooling 
lines may be used to constrain physical parameters and possibly 
the age of the starburst. These ratios are sensitive to the 
physical and chemical conditions (density, temperature, and abundance), 
hence provide an opportunity to model the physical state of the 
neutral gas. In addition, unlike optical atomic line tracers, 
these FIR lines are relatively insensitive to extinction. 
A close examination of these lines emitted in M 82 may provide 
a template for future comparisons to infrared-bright, dust 
obscured starburst galaxies like M 82, including those at high-$z$. 

In order to provide a useful indication on the starburst age(s),
it is desirable to make comparisons with multiple transitions 
for various molecules and atoms. However, meaningful comparisons 
can be made only for regions where observations refer to the same 
beam size. The diagnostic tracers used in this paper are molecular 
$^{12}$CO, its isotope $^{13}$CO, HCN, HCO$^+$, and atomic C, O, 
and C$^+$. The low-$J$ $^{12}$CO lines are easily excited at 
relatively low densities and temperature, and are found essentially 
in every molecular gas cloud, and so they are good diagnostic tools 
for total molecular H$_2$ content, diffuse gas conditions, and star 
formation history. The less abundant $^{13}$CO isotope has a much 
lower optical depth, and the line ratios between optically thin 
transitions in $^{13}$CO are more reliable probes of the total gas 
content than $^{12}$CO. The CO molecule is not considered a good 
tracer of dense and highly excited gas that is directly involved 
in starburst (i.e. earlier phase of star formation). However, 
molecular HCN and HCO$^+$ lines are more sensitive to dense gas 
(i.e. pre- or post-birth of stars) owing to their higher critical 
densities than CO. The atomic C, O, and C$^+$ fine structure lines are 
excellent probes of the PDRs in starburst regions, and their line 
ratios can be used for diagnosing the conditions of the associated 
FUV flux and gas density, as well as for indicating the ages of 
the later stages of starbursts. 

In this paper, we use the observations of molecular and atomic gas 
in the central 1 kpc of M 82 described above. The molecular data for 
the center 1 kpc (along the major axis of the disk) are taken from 
Weiss et al. (2005). The molecular data for the center 680 pc 
are taken from Mao et al. (2000). The atomic data 
\citep[][and references therein]{neg01} to be 
used in our ratio-ratio diagram analysis, which are obtained from a 
larger area ($\sim$ 1.2 kpc) than for the molecular data. Note that 
the atomic C data are not included in our later ratio-ratio analysis, 
because the two [C I] line data correspond to different beam sizes 
\citep{stu97}. These molecular data will be used in comparisons with 
our models.   

\subsubsection{Model Fit to the $^{12}$CO Line Spectral Energy Distribution} \label{lsed}

We consider first a model involving a single instantaneous starburst, 
and later consider whether extended starbursts could also provide an 
adequate fit.

\begin{center}
{\it Part I: Instantaneous Starburst Model} 
\end{center}

Our initial starburst model involves a single event in which all of 
the stars/clusters are formed simultaneously and instantaneously in 
the center of GMCs, associated with a unique age $t$ and a star cluster 
mass $M_{\ast}$ (corresponding to a GMC mass $M_{GMC}$ = 4$M_{\ast}$ 
at the birth time). An instantaneous model, though physically unrealistic, 
is an acceptable representation of the SED if the duration of the star 
forming event is short compared to the age of the starburst. The 
intent is to derive these two parameters by fitting to the data. 
By extension, the total mass of H$_2$ swept-up in the ISM at any age 
is also determined.

The method used for fitting is the well-known numerical chi-squared 
($\chi^2$) procedure. The  number of data points is six for the $^{12}$CO 
data, and the number of free parameters (in this case) is two. The quantity 
$S^J_{model}$ represents the corresponding model to be fitted, 
$S^J_{model}$($f$, $t$) = $f$ $S^J_{temp}$ ($t$), where $S^J_{temp}$($t$) 
is the model template line SED at age $t$, as given in Fig.~\ref{lsed_co1} 
(see \S~\ref{ss}), corresponding  to a model template GMC mass $M^{GMC}_{temp}$, 
a model template cluster mass $M^{\ast}_{temp}$, and a model template swept-up 
mass by the shells $M^{sh}_{temp}$. The values for these parameters are 
$M^{GMC}_{temp}$ = 1.69 $\times$ 10$^7$ M$_{\odot}$, and 
$M^{\ast}_{temp}$ = 4.2 $\times$ 10$^6$ M$_{\odot}$. 
These initial masses correspond to the 127 clusters included in 
Table~\ref{tbl1} (see \S~\ref{esbm}). The ratio of stellar cluster 
to GMC mass is 0.25 according to the assumed SFE. The adjustable 
dimensionless parameter $f$ is introduced to control the amplitude 
of the model line SED (and hence the total cluster mass), and 
the age parameter $t$ controls its shape and slope. These are 
simultaneously adjusted to provide the best fit corresponding 
to the minimum $\chi^2_{\nu}$. By assumption, the line fluxes 
$S^J_{temp}$($t$) are summed over the contributions of all clusters 
and GMCs, so that the best fit GMC mass $M_{GMC}$, cluster mass 
$M_{\ast}$, and the shell swept-up mass $M_{sh}$ are determined 
from the corresponding best fit value of the parameter $f$ by 
the relations,

\begin{eqnarray}
M_{GMC} & = & f M^{GMC}_{temp}, \\
M_{\ast} & = & f M^{\ast}_{temp}, \\
M_{sh} & = & f M^{sh}_{temp}
\end{eqnarray} 

\begin{figure}
\epsscale{1.0}
\plotone{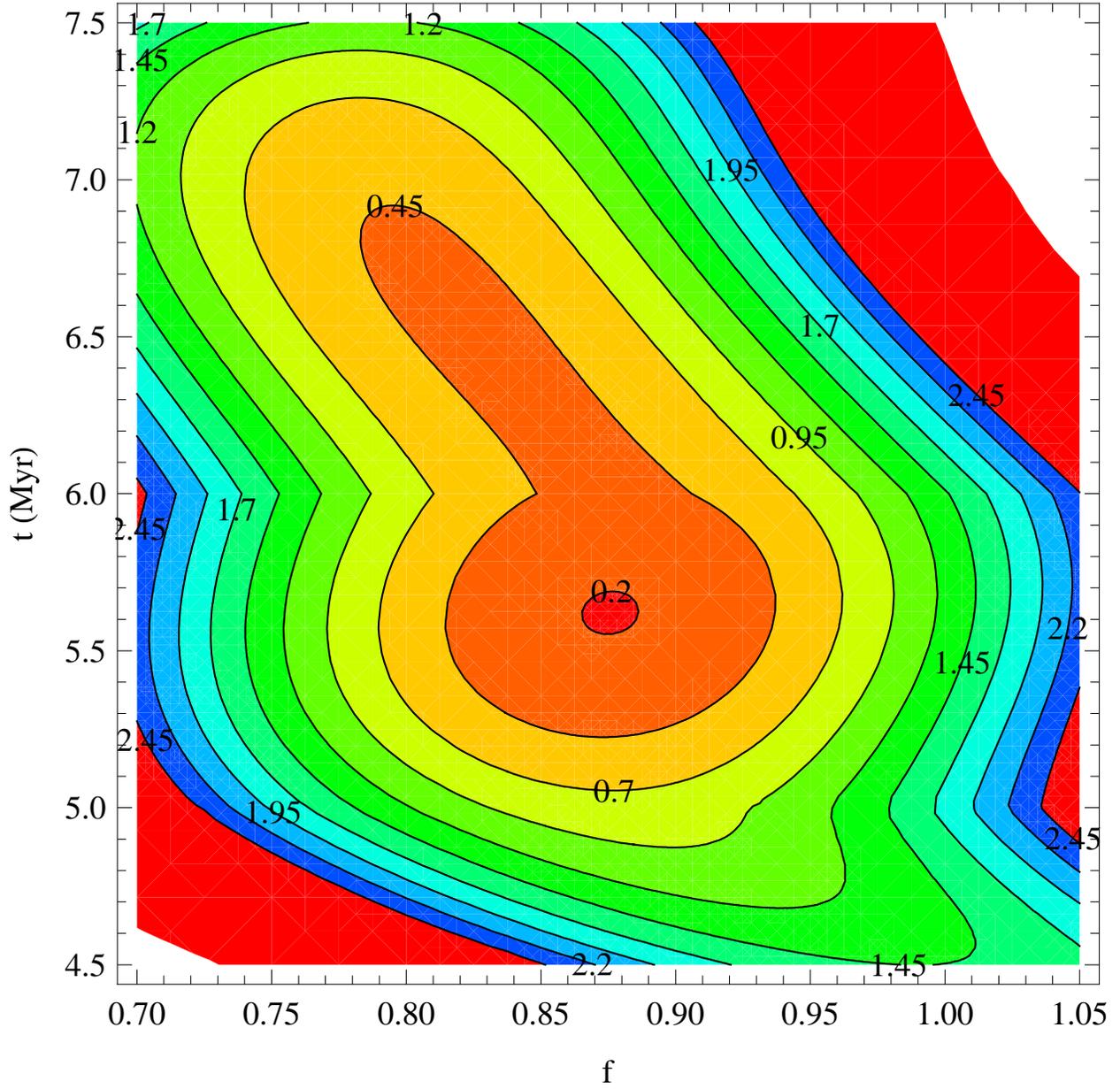}
\caption{A contour plot of $\chi^2$ values as a function of mass coefficient 
$f$ and burst age $t$ at the {\it post-SN} phase.
\label{chi2_co}}
\end{figure}

We calculate $\chi^2_{\nu}$ for a range of $t$ and $f$. A minimum 
$\chi^2_{\nu}$ value is obtained with a standard error estimation (i.e. 
the traditional likelihood method from using an inverse Hessian matrix 
or covariance matrix). In the results there were found to be two minima 
with acceptable values of $\chi^2_{\nu}$(min), one for the {\it Winds}
phase and the other for the {\it post-SN} phase. The chi-squared 
contour surrounding the minimum for the {\it post-SN} phase is shown 
in Fig.~\ref{chi2_co}, where the contours are 
$\chi^2_{\nu}$ = $\chi^2_{\nu}$(min) + $i$ / ($N$ - $p$), and 
$i$ = 1, 2, 3, $\ldots$, corresponding to contour intervals of 1$\sigma$. 
The existence of a numerically acceptable solution for each phase signifies 
that there exist conditions in the molecular clouds of the {\it Winds} phase 
which are similar to those found in the compressed shells associated with 
the {\it post-SN} phase. However, only the {\it post-SN} phase solution
is acceptable physically, since an age of 0.07 Myr derived for the 
{\it Winds} phase is implausibly small for a variety of reasons. 
For example, it is impossibly short compared to the dynamical time 
for the region (a few Myr) which would control the duration of the 
starburst. Such a small age might be barely plausible for an individual 
shell, but not for the molecular gas occupying this entire region. 

\begin{figure}
\epsscale{1.0}
\plotone{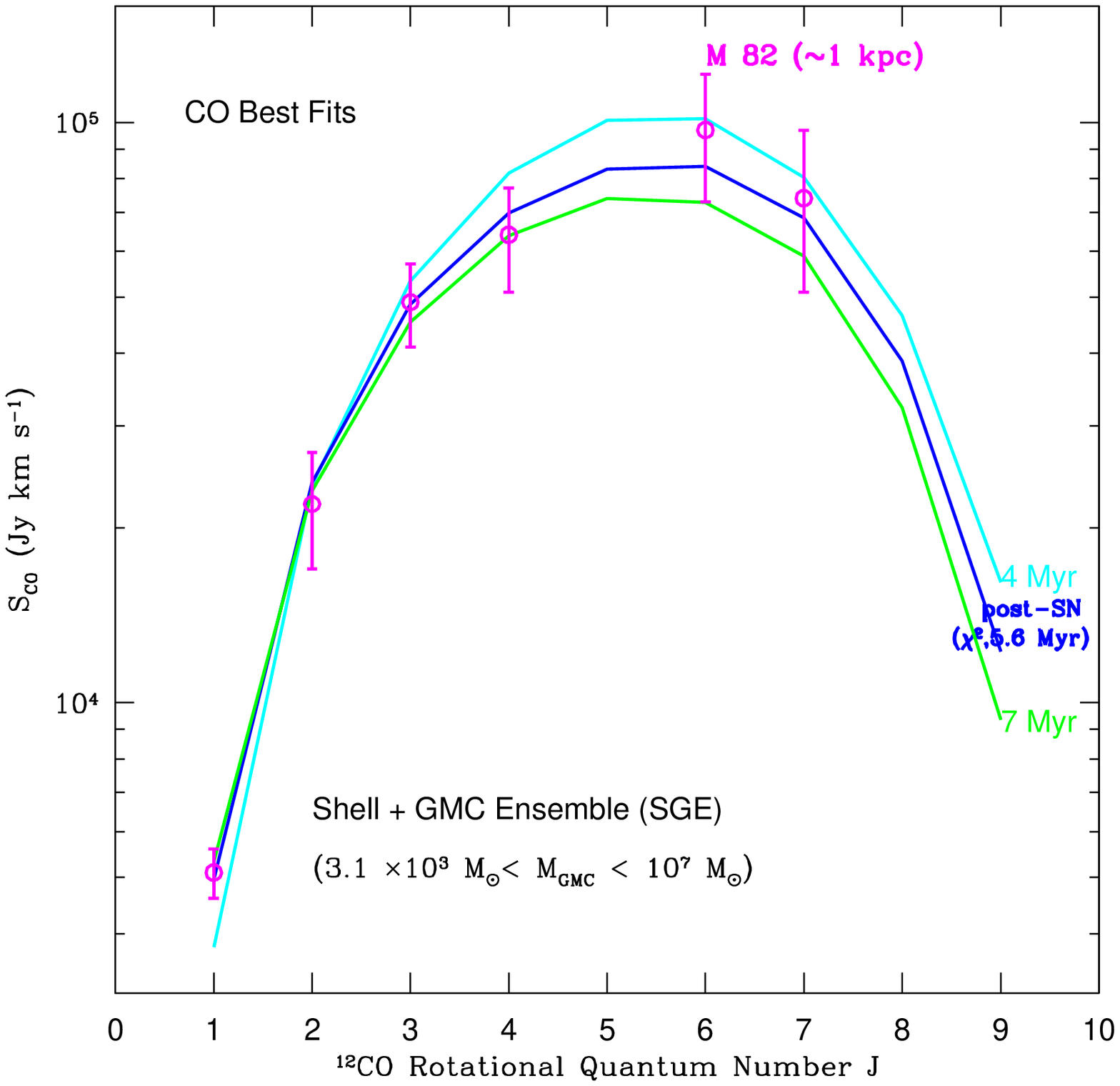}
\caption{A $\chi^2$ fit of an instantaneous starburst model to the $^{12}$CO 
line SED for the central 1 kpc disk region of M 82. The blue curve is the 
best age at 5.6 Myr for the {\it post-SN} phase, the cyan curve is the line 
SED at 4 Myr, the green curve is the line SED at 7 Myr, and the observed 
data are indicated by magenta open circles with error bars \citep{wei05}.
\label{cofit}}
\end{figure}

The best fit line SEDs and $^{12}$CO line SEDs at 4 and 7 Myr are shown 
superposed on the data in Fig.~\ref{cofit}. The initial stellar mass and 
GMC mass are relatively small, but the impact on the surrounding ISM 
is significant. The total H$_2$ mass swept up by the shells is 
$\sim$ 2.0 $\pm$ 0.1 $\times$ 10$^8$ M$_{\odot}$ at the best fit age 
of 5.6 Myr. This predicted value is in good agreement with the total gas 
mass within the central 1 kpc region obtained by other studies (a few 
10$^8$ M$_{\odot}$) \citep[e.g.][]{rie80, mao00}. The model total stellar 
luminosity at this best fit is $\sim$ 1.4 $\times$ 10$^9$ L$_{\odot}$ 
using information from {\it Starburst99} based on our instantaneous 
starburst model. 

In order to investigate how sensitive the results are to the assumed 
initial upper mass limit of the cluster spectrum (and corresponding 
GMC mass spectrum), we repeated the above analysis with revised upper 
mass limits of both 7.5 $\times$ 10$^5$ M$_{\odot}$ and 
2.5 $\times$ 10$^5$ M$_{\odot}$ for the stellar spectrum, and 
corresponding GMC upper mass limits of 3 $\times$ 10$^6$ M$_{\odot}$
and 10$^6$ M$_{\odot}$ respectively. For the first case, we find 
$t$ = 5.0 $\pm$ 0.4 Myr, $f$ = 1.8 $\pm$ 0.2 with $\chi^2_{\nu}$ = 0.9,
and for the second case an unacceptable fit with a $\chi^2_{\nu}$ = 33.
Thus, a comparable solution may be found with a choice of a slightly 
lower upper mass cutoff, but no acceptable solutions are found with 
values reduced by a factor of 10 or more in the upper cutoff of 
the cluster mass spectrum. We conclude that the model can provide 
acceptable fits to the data only if the dominant initiating starburst 
clusters are massive, at least 5 $\times$ 10$^5$ M$_{\odot}$.

Our models show that the H$_2$ density of the shells at the best 
fitted age 5.6 Myr is between 10$^3$ and 10$^4$ cm$^{-3}$, and the 
gas temperature is $\sim$ 50 - 100 K. These values are comparable 
with the two-component LVG predictions \citep{wei05}. The evolution 
of CO abundance as a function of $A_V$ is illustrated in Fig.~\ref{x2Av1}.
The CO abundances in massive shells (M6 - M7) are above 10$^{-5}$ 
with respect to the total H density, providing most of the CO emission. 

\begin{center}
{\it Part II: Extended Starburst Model} 
\end{center}

The foregoing discussion and results assume an instantaneous starburst 
with the result that our CO best fit model has an age of 5.6 Myr. 
The question naturally arises whether a model with a period of more 
continuous star formation would also provide a satisfactory solution. 
One can anticipate that the answer might be yes, if the best fit line 
SED were to be roughly equally represented by the SED of an outburst 
at one epoch, or an average SED over some time period roughly centered on, 
and symmetrically distributed about this epoch. Fortunately, it is 
straight forward to test this hypothesis since the SED for a smoothly 
varying star formation model may be constructed from a superposition 
of instantaneous bursts at different times.

We are thus led to consider the extreme case of a uniform star formation 
rate (or SFR) occurring between an epoch 10 Myr ago and the present time. 
This starting point of the event may be considered appropriate because 
the most massive shells from even earlier epochs would now be large enough 
to exceed the thickness of the nuclear disk and thus their emission 
would begin to fall outside the region modeled. Fortuitously, this period 
is also almost symmetrically distributed about the epoch for the best fit 
instantaneous model. 

We do not discuss the procedure in detail here, since the analysis proceeds 
as before, but with only one parameter, namely the star formation rate 
over the past 10 Myr. The SED employed is then an integral of the CO 
line SED profiles over a time period of 10 Myr. The result is 
that an acceptable fit (minimum $\chi^2_{\nu}$ = 0.78) can be found for 
a continuous $SFR$ = 0.5 $\pm$ 0.05 M$_{\odot}$ yr$^{-1}$. The total stellar 
mass produced during this period is (5.0 $\pm$ 0.5) $\times$ 10$^6$ M$_{\odot}$, 
which is, not surprisingly, close to the total mass (i.e. 4.3 $\times$ 10$^6$ 
M$_{\odot}$) required for the single epoch model. The conclusion is that a 
uniform $SFR$ over the past 10 Myr also produces a satisfactory fit to the 
$^{12}$CO data. In addition, it may be plausibly inferred that a variety 
of star formation histories would work, provided the $SFR$ rate profile 
is more or less symmetrically distributed about the epoch of 5.6 Myr. 

The implication of this result is that in terms of the agreement between 
the model and the data, the star formation need not be instantaneous, 
or even sharply peaked at 5.6 Myr. However this epoch nevertheless represents 
a unique point of time associated with the history of star formation in M 82 
since it would emerge from various representations of the star formation 
profile. If there were an instantaneous starburst in M 82, then its 
age is approximately 5 - 6 Myr (plus or minus a factor of about 2). However, 
the SED is also consistent with continuous star formation rate of 
0.5 M$_{\odot}$ yr$^{-1}$ over at least the last 10 Myr. The factor of 
2 is our estimate of the uncertainties introduced by the uncertain density 
of the ambient interstellar medium in the model, as well as our assumption 
that the shells do not stall.

\subsubsection{Molecular and Atomic Line Ratio Diagrams} \label{lratio}

The line intensity/flux ratio-ratio diagram can be another diagnostic tool 
for studying the gas excitation conditions and properties, as well as
their relations to starburst evolution, especially when there are not 
enough data points available for the type of model fitting discussed 
in the previous section. Since the line ratio is independent of the 
total gas mass in the measured region, the ratio-ratio diagram cannot 
be used to provide an estimate of the total gas mass directly. However, 
once the age $t$ is obtained, we can scale the template line flux spectrum to
the flux observed, and calculate the model value for the swept-up H$_2$ mass.

\begin{center}
{\it \large $^{12}$CO and Its Isotope $^{13}$CO}
\end{center}

\begin{figure}
\epsscale{1.0}
\plotone{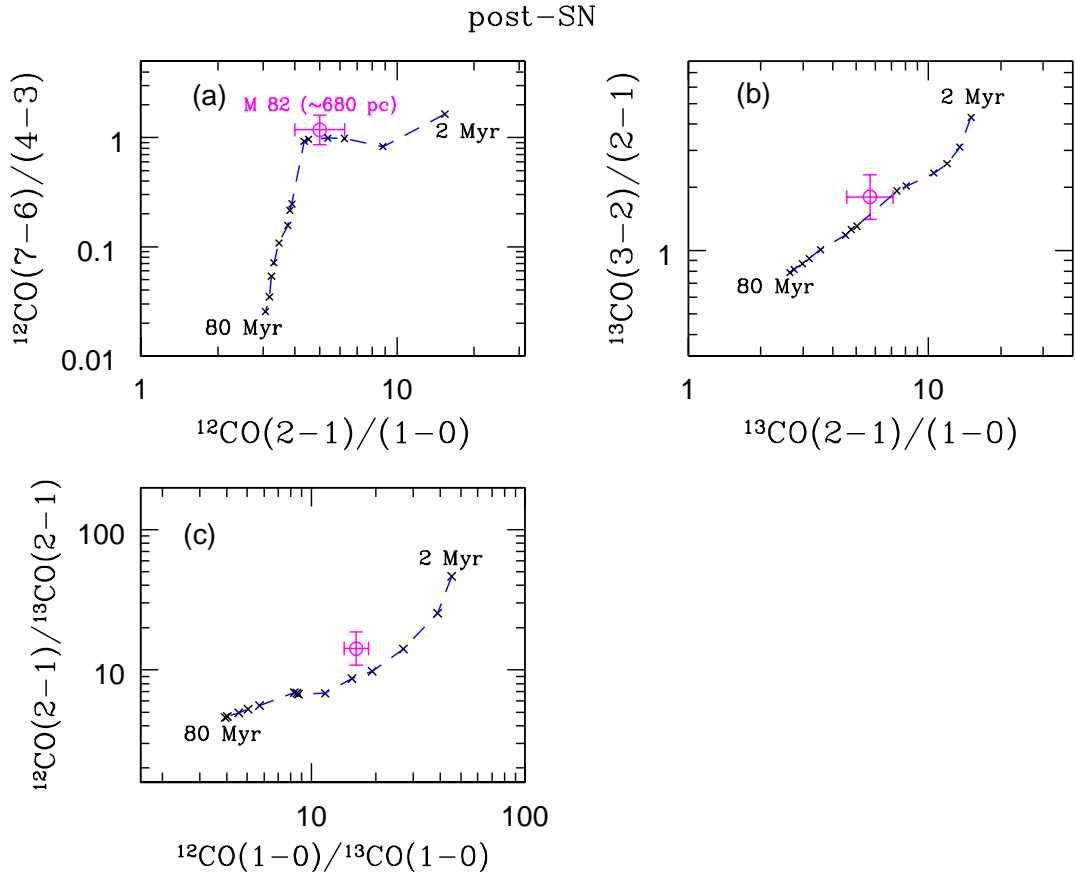}
\caption{The ratio-ratio diagrams of molecular $^{12}$CO and $^{13}$CO 
line intensities (in units of Jy km s$^{-1}$). Model results for a shell 
ensemble are indicated by the crosses connected with blue dashed lines. 
The age sequence is 2, 3, 4, 5, 6, 7, 8, 9, 10, 20, 30, 40, 60, and 80 Myr. 
The magenta filled circles with error bars are the observed data 
(22$^{\arcsec}$ resolution data for the center of M 82 from Table 1 of 
Mao et al. 2000; the errors include systematic uncertainties).
\label{rr_co1}}
\end{figure}

Fig.~\ref{rr_co1} illustrates the ratio-ratio diagrams for different 
transitions involving $^{12}$CO and $^{13}$CO predicted by our 
model ({\it post-SN}), and a comparison with the observations. 
The observed data refer to the center of M 82 with a beam-width 
of $22^{\arcsec}$ (see Table 1 in Mao et al. 2000). All line 
brightnesses are compared in units of Jy km s$^{-1}$. The isotope 
abundance ratio [$^{12}$CO]/[$^{13}$CO] of 55 is adopted for 
the ensemble modeling. In the plots, we include the systematic 
uncertainties (31\% for $^{12}$CO(7-6), 20\% for $^{12}$CO(4-3), 
16\% for $^{12}$CO(3-2), 23\% for $^{12}$CO(2-1), 10\% for 
$^{12}$CO(1-0)) into the line ratio error estimations (i.e. sizes 
of error bars). In plot (a) the model ratios of $^{12}$CO(7-6)/(4-3) 
versus (2-1)/(1-0) match nicely (as expected) with the observations 
at age 5 - 6 Myr for the center 680 pc region. It is similar 
to the age derived from the chi-squared fit to the $^{12}$CO line 
SED in the center 1 kpc region \citep {wei05} even though the 
angular size of the region is different. This is expected, since 
$\sim$ 65 - 80\% of the $^{12}$CO emission from the inner 1 kpc 
disk originates from the central 680 pc starburst regions. 
Plots (b) and (c) show a poor match between our model line ratios 
of $^{13}$CO(3-2)/(2-1) versus (2-1)/(1-0) and 
$^{12}$CO(2-1)/$^{13}$CO(2-1) versus $^{12}$CO(1-0)/$^{13}$CO(1-0) 
(blue-dashed curves) and the observed data. The closest match within 
the observed uncertainties is 7 - 8 Myr for plot (b), 5 - 6 Myr for  
plot (c), where the latter is in a fair agreement with $^{12}$CO best 
fitted age. 

Our model fails to produce the right ratios for lines involving $^{13}$CO. 
If the choice of the isotope abundance ratio is to be considered as 
the reason for such poor fit, adopting a different isotope abundance 
ratio (55 is used in this paper) can affect the result in plot (c) but 
not that in plot (b). Mao et al. (2000) indicated that their $^{13}$CO(2-1) 
values should be considered with caution, due to uncertainty of convolving  
a smaller beam (13$^{\arcsec}$) to a larger beam size (22$^{\arcsec}$).  
If we assume that the best match age for plot (b) and (c) should 
be between 5 and 6 Myr, and if we assume that an erroneous value for 
the $^{13}$CO(2-1) model flux is the reason for lower ratios seen 
in plot (b) and (c), we estimate that this value is underestimated 
by a factor of 1.5.  Another factor that contributes to the poor 
match between our model results and the data is that the line ratios 
involving $^{13}$CO may be particularly sensitive to optical depth 
if the intensities are not optically thick. 

We also investigated the effect of reducing the upper mass limit to
the GMC mass spectrum as was done for the $^{12}$CO model, and found that 
changing the upper mass limit to the GMC mass spectrum has no effect 
in resolving this problem.            
 
\begin{center}
{\it \large Atomic O and C$^+$}
\end{center}

\begin{figure}
\epsscale{1.0}
\plotone{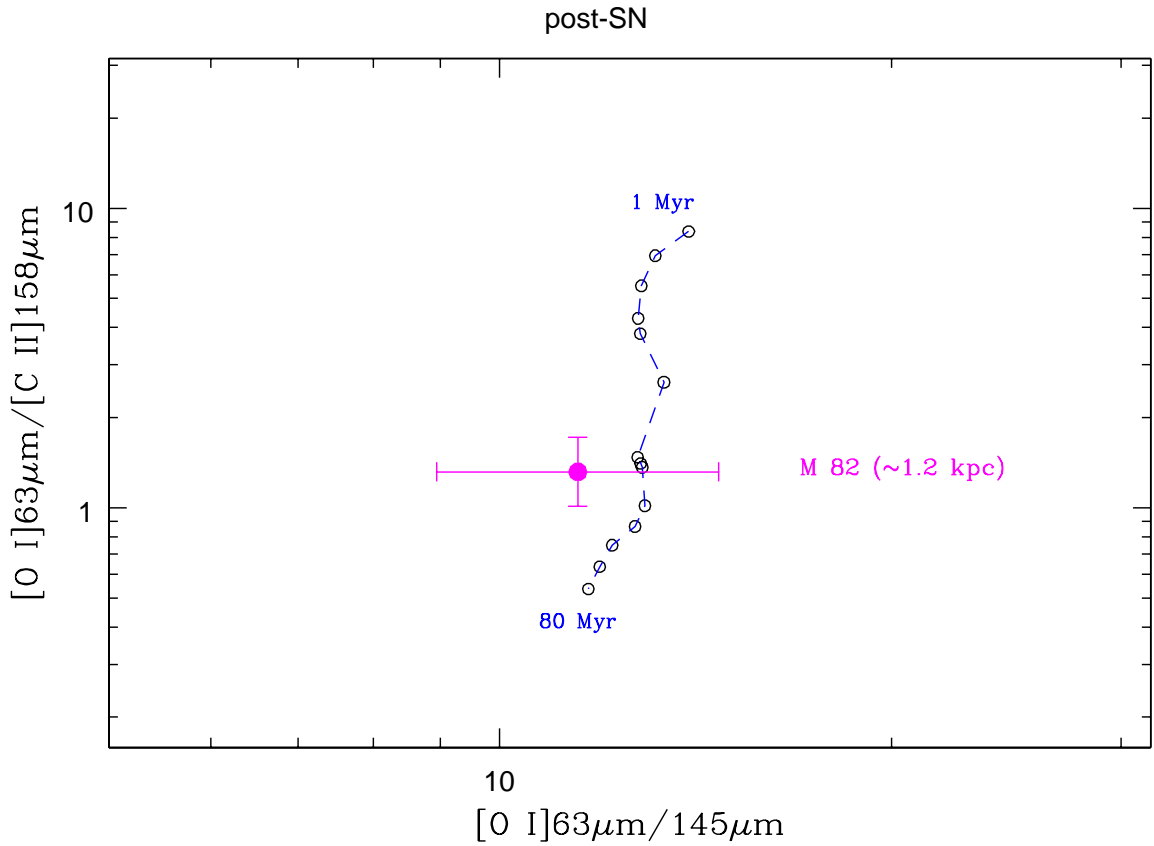}
\caption{The ratio-ratio diagram of atomic fine structure line fluxes (in units 
of W m$^{-2}$). The models are indicated by open circles connected with a 
blue-dashed curve for {\it post-SN} phase. The age sequence is 1, 2, 3, 4, 
5, 7, 8, 9, 10, 20, 30, 40, 60, and 80 Myr for the {\it post-SN} phase. 
The filled circle with error bars show the observed data for the center 
1.2 kpc of M 82 \citep{neg01}. 
\label{rr2t_atom}}
\end{figure}

Fig.~\ref{rr2t_atom} shows the model ratio-ratio diagram for
[O I]63$\mu$m/[C II]158$\mu$m versus [O I]63$\mu$m/[O I]145$\mu$m, 
and a comparison with the observations of these atomic lines from 
the central 1.2 kpc region in M 82 \citep{neg01}. All line fluxes 
are compared in units of W m$^{-2}$. The model [C II]158$\mu$m 
line flux may be underestimated, since we ignore the line emission 
that arises from the H II region. A good match between our model 
and the observation is obtained with age $t$ $\sim$ 10 Myr. 
The age predicted from atomic data is older than the age 
($\sim$ 5 - 6 Myr) derived from our $^{12}$CO line SED analysis. 
This may be because the atomic line data are based on a 
80$^{\arcsec}$ $\times$ 80$^{\prime\prime}$ beam area whereas the 
$^{12}$CO line data pertain only to the 60$^{\arcsec}$ $\times$ 18$^{\arcsec}$ 
beam area. We suggest that these two ages may be a result of 
sampling different regions. More discussion of this possibility
will be given in \S~\ref{disc_cen}. 

The ranges of gas conditions for the model shells at 10 Myr are 
$G_0$ $\sim$ 350 - 1.4 $\times$ 10$^4$, $n$(H$_2$) $\sim$ 10$^2$ - 
2.4 $\times$ 10$^3$ cm$^{-3}$, and $T_{gas}$ $>$ 20 K. The gas 
conditions derived from our atomic models for this sampling region 
are comparable with the study by Colbert et. al. (1999) 
($G_0$ = 630, $n$ = 2.0 $\times$ 10$^3$ cm$^{-3}$), but the age 
is greater than that (3 - 5 Myr) derived by Colbert et al.. 

The total molecular gas swept up into shells cannot be obtained directly 
from the ratio-ratio diagram. However we can obtain this from the ratio 
of observed [O I]63 $\mu$m line flux to the model template [O I]63 $\mu$m 
flux at age 10 Myr, i.e. $f$ = $S_{obs}$ / $S^{temp}_{model}$ = 0.84 
, where $S_{obs}$ = 169 $\times$ 10$^{-15}$ W m$^{-2}$ and $S^{temp}_{model}$ 
= 202 $\times$ 10$^{-15}$ W m$^{-2}$. We compute the total H$_2$ gas 
in the measured 80$^{\arcsec}$ region by multiplying $f$ = 0.84 by 
the model template H$_2$ mass $M^{temp}_{model}$. Hence, the result 
is $M$(H$_2$) at age 10 Myr is $\sim$ 3.4 $\times$ 10$^8$ M$_{\odot}$.

\subsection{Discussion} \label{m82disc}

\subsubsection{An Expanding Supershell Associated with SNR 41.9+58} \label{disc_ss}

The very good agreement between our supershell kinematic model and 
the observations is consistent with the hypothesis that this 
expanding supershell is created by strong mechanical winds from 
a young star cluster with a total mass of about 
2.5 $\times$ 10$^6$ M$_{\odot}$ which formed at its center about 
0.8 Myr ago. This agreement also suggests that the set of models 
we have put forward in this paper may be used to interpret other 
shells in M 82 or shells in other starburst galaxies. Although 
like any other model, the result depends to some degree on the 
set of initial conditions and assumptions that we selected for 
our models. The reliability of the age and mass for this supershell 
derived from our kinematic study needs to be further examined in near 
future when high resolution maps of multiple transitions of CO emission 
in this shell are available to compare with our model. Meanwhile
we relate our model CO line ratio SED (i.e. $I_{CO}$/$I_{CO(1-0)}$
as a function of $J$) for the SNR 41.9 + 58 (i.e. M7) supershell 
at age 0.8 Myr to the corresponding line SED for the surrounding 
gas in the inner 1 kpc starburst region of M 82. We show that 
the emission in the M7 supershell exhibits a higher degree 
of excitation than the surrounding emission (see Fig.~\ref{ssco}). 
 
There are a number of issues arising from the supershell study. They 
are as follows: (1) it is interesting to ask whether our results are 
consistent with a possible physical association between the supershell 
and the bright SNR 41.9 + 58 near its center. If the bright SNR were 
within or near the SSC, there may not be sufficient gas remaining 
to form an SNR after the action of the winds from the cluster; 
and (2) the SSC responsible for the formation of the supershell might 
also have provided the stellar mass for the several hundred solar 
mass black hole detected by Chandra X-ray observations near its center 
\citep[e.g.][]{dew06}. Theories for the formation of this black hole 
include the collapse of a {\it hyperstar} formed by the coalescence 
of many normal stars, or the direct merger of stellar mass black holes 
\citep[e.g.][]{kau05}. The SSC is adequately endowed with sufficient 
mass since there would have been 1,700 O stars, each with mass greater 
than or equal to about 40 M$_{\odot}$ \citep{yao06}.   

\subsubsection{Central Starburst Region} \label{disc_cen}

\begin{center}
{\it \large Age of Recent Starburst and Star Formation History} 
\end{center}

Given the complexity of M 82, a full understanding of star formation 
epochs requires various diagnostic tools to trace different ISM
components in starburst regions. Especially since both optical and 
near- to mid-infrared emissions are subject to higher extinction in
dusty media, the selection of SSCs may be biased toward either 
younger or older age as mentioned before. Since the ISM is nearly 
transparent to FIR/sub-mm/mm emission, the analysis in this paper, 
employing atoms and molecules emitting in this range, forms a useful 
complementary investigation to those already mentioned. Using our 
evolving starburst model for neutral gas media, we have been able 
to probe the recent star formation history of M 82 throughout the 
entire volume of the central starburst region. 

The starburst ages derived from optical and infrared spectra are 
5 - 6 Myr, 10 - 25 Myr, and 30 - 100 Myr. The age derived from 
our analysis of CO line SEDs and ratio-ratio diagrams is 5 - 6 
Myr for the central 1 kpc $\times$ 280 pc rectangular regions; 
although the region used is 1 kpc, about 70\% is concentrated 
toward more central regions ($\sim$ 350 pc). The age derived 
from our atomic data is slightly older, i.e. 10 Myr for a larger 
area ($\sim$ 1.2 kpc). We suggest that these two ages may be 
a result of sampling different regions as mentioned earlier. 
It is unclear from our analysis whether these two ages refer to 
the same period of star forming activity or to two spatially 
separated independent bursts. A more sophisticated model and 
more data are needed to clarify the picture.

The burst ages derived from our model are similar to the results 
found in the aforementioned studies by F\"{o}rster-Schreiber et al. 
(2003) and Efstathiou et al. (2000). However, for the atomic data 
there is a discrepancy between our result (10 Myr) and the study 
by Colbert et al. (1999) (3 - 5 Myr) using a similar set of data 
(by Negishi et al. 2001). Nevertheless, our derived gas conditions 
for the shells at 10 Myr ($G_0$ $\sim$ 350 - 1.4 $\times$ 10$^4$, 
$n$(H$_2$) $\sim$ 10$^2$ - 2.4 $\times$ 10$^3$ cm$^{-3}$, and 
$T_{gas}$ $>$ 20 K) are similar to those derived by Colbert et al.. 
The age discrepancy may be caused by differences in the choice 
of models, for example, Colbert et al. used the CLOUDY PDR model 
to compute the atomic line fluxes, while we use the UCL\_PDR model 
and a non-LTE radiative transfer model to compute the line fluxes.

\begin{center}
{\it \large Molecular Gas Properties} 
\end{center}

Our evolving shell models yield familiar values for the gas density, 
temperature, and structure scales compared to those measured in the 
center of M 82 \citep[e.g.][]{las63, rie89, stu97, saf00, mao00, neg01, war03}. 
For an extended starburst scenario ($SFR$ = 0.5 $\pm$ 0.05 M$_{\odot}$ yr$^{-1}$
as discussed in \S~\ref{lsed}), the shell densities are in the range 
10$^2$ - 10$^6$ cm$^{-3}$, and the gas temperatures are in the range 
20 K to 1000 K across the shell for various shells. The total H$_2$ mass 
swept up by the shells within the inner 1 kpc 
($\sim$ 2.0 $\pm$ 0.1 $\times$ 10$^8$ M$_{\odot}$) and 
1.2 kpc ($\sim$ 3.4 $\pm$ 0.3 $\times$ 10$^8$ M$_{\odot}$) detection 
regions are compatible with those derived from the CO luminosity using the 
CO-to-H$_2$ conversion factor \citep{wil92, mao00, wal02}. It is also 
comparable with the total ambient gas mass in our model. Hence, the 
picture suggested is that of a {\it porous} neutral ISM in the central 
star-forming region of M 82, a product of evolving shells. In reality, 
many or most shells are probably in the form of fragments, small 
cloud clumps, sheets, or partial and full circular arcs 
\citep[e.g.][and references therein]{lok87, yao06}.

\begin{center}
{\it \large Molecular Ring Formation Mechanism} 
\end{center}

Although different stages of starburst evolution are applicable to 
different central regions of M 82, the shell sizes and the physical 
conditions of the gas within the rings (diameter $\sim$ 300 - 600 pc) 
predicted by our model are similar to what is expected from models 
involving expanding shells from a central starburst such as those 
proposed by Carlstrom \& Kronberg (1991). Their hypothesis is that 
molecular rings in M 82 are a result of compressed gas in a starburst 
region. This hypothesis is supported by the observations of the 
geometrical structure of the CO line emission and continuum emission, 
as well as the discovery of supershells that have not yet had time 
to break out of the galactic plane. However, the conclusion drawn 
from the shell size and average gas conditions in the inner 1 kpc 
region is only suggestive, since our model does not handle the 
physical distribution of molecular gas in the center of M 82. It is 
also important to realize that the foregoing interpretation of the 
lobes as a ring is not unique. A number of authors have argued that 
the molecular rings are a product of Linblad resonance instabilities 
associated with the gravitational effects of the bar 
\citep[e.g.][]{sal96, wil00}.

\subsubsection{Limitations of Our Model and Their Impacts} \label{limit}

We have demonstrated that the kinematic and FIR/sub-mm/mm emission 
properties of individual expanding shells and star-forming regions 
in a starburst galaxy like M 82 can be understood by following 
the evolution of individual massive super star clusters or an 
ensemble of such clusters surrounded by compressed shells and GMCs. 
It is an important piece of complementary work to the existing 
optical and infrared studies, and it helps us to obtain a more 
complete and or accurate picture of star formation episodes in 
the center of M 82. 

However, our model also has a number of caveats, limitations, and 
potential sources of systematic error. Here is the important list:

(1) We have neglected throughout the effects of the ambient pressure 
in slowing down and perhaps stalling the shells. This applies to both 
{\it Winds} and {\it post-SN} phases. We recall that the shells 
will stall when their expansion velocities decrease sufficiently that 
they are approximately equivalent to the sound speed ($P$/$\rho$)$^\frac{1}{2}$ 
of the external medium. To estimate the effects of this pressure, 
we can thus compute the sound speed associated with estimates 
of the pressure and compare this with the shell speeds.  
We compute the total pressure $P_{cloud}$ inside the cloud, 
assuming it is in virial equilibrium from the following 
equations \citep{mak87}, 
 
\begin{eqnarray}
P_{cloud} & = & P_{ext} + P_{intern}, \\
P_{ext} & = & 2n_0 k T, \\
P_{intern} & = & 0.5 G \Sigma^2 
\end{eqnarray}

where $P_{exter}$ is the external pressure, $P_{intern}$ is the
internal pressure, $n_0$ is the cloud H$_2$ density, $k$ is the Boltzmann 
constant (1.38 $\times$ 10$^{-16}$ erg K$^{-1}$), $G$ is the gravitational 
constant (6.67 $\times$ 10$^{-8}$ cm$^3$ g$^{-1}$ s$^{-2}$), 
and $\Sigma$ = $M_{GMC}$ / ($\pi$ $R^2_{GMC}$). The sound speed 
in a GMC can be calculated from the equation,

\begin{equation}
c_{cloud} = \Big(\frac{P_{cloud}}{\rho_{cloud}}\Big)^{\frac{1}{2}}
\end{equation}

We obtain a sound speed of $c_{cloud}$ = 19 km s$^{-1}$ in a M7 cloud
($n$(H$_2$) = 300 cm$^{-3}$, $M_{GMC}$ = 10$^7$ M$_{\odot}$, and 
$R_{GMC}$ = 47 pc), assuming $P_{ext}$/$k$ = 10$^7$ K cm$^{-3}$ 
in starburst regions of M 82 \citep{sil07}, where $k$ is the Boltzmann 
constant. We can further combine Equations (16) through (19) with 
the cloud relations Equations (7) and (8) to furthermore yield 
the sound speed in any given GMC,

\begin{equation}
c_{cloud} = 19 \ km s^{-1} \Big(\frac{M_{GMC}}{10^7 M_{\odot}}\Big)^{\frac{1}{4}}
\end{equation}

For the external ambient medium, we use the aforementioned external 
pressure to obtain the sound speed for the ISM ($n_{ism}$ = 30 cm$^{-3}$),

\begin{equation}
c_{ism} = 40 \ km s^{-1}
\end{equation}

The comparison between sound speeds inside the clouds (Equation (20)) 
and shell expansion velocities for the {\it Winds} phase indicates 
that shells from cloud masses above 10$^5$ M$_{\odot}$ would not 
be trapped, and those equal or below this mass would be stalled
if the effects of cloud pressure were included. 

We also compare the sound speeds in the ISM (40 km s$^{-1}$)
with shell expansion velocities for the {\it post-SN} phase.
For example, for the shells associated with the three most massive 
GMCs, namely M7, 3M6, and M6 in our model ensemble, the shells have 
radii of 220, 180, and 130 pc at the best fit age 5.6 Myr without
ambient ISM pressure, respectively. But the stall radii and ages 
for these three shells are 112 pc at 2 Myr, 70 pc at 1.2 Myr, 
and 42 pc at 0.8 Myr, respectively. Thus, without the inclusion 
of the effects of pressure it may be said that the shell radii 
at the time of observation are probably overestimated by more than 
a factor of two compared to the stall values when pressure is included. 
Since the swept-up mass by the shell is proportional to the $R^3_s$, 
an overestimate by a factor of two in shell radius would yield
a factor of eight in the total swept-up mass for a given GMC/SC 
mass. This may help to understand the shortfall in IR 
luminosity predicted by our starburst model (see point (3)
for detailed discussion).

Another issue worth mentioning is that the confining pressure will 
vary greatly with location in the galaxy, especially between the 
center and the edges of the disk where some of the observed supershells 
are located. For example, as we mentioned earlier in this chapter, 
observations have detected an expanding supershell centered around 
the bright SNR 41.9 + 58 in both molecular line and radio continuum 
\citep[e.g.][]{wei99, wil99}. This supershell has a diameter of 
$\sim$ 130 pc, an expansion velocity of $\sim$ 45 km s$^{-1}$, 
and a mass of $\sim$ 8 $\times$ 10$^6$ M$_{\odot}$. 
If $P_{ext}$/$k$ = 10$^7$ K cm$^{-3}$ were the relevant external 
pressure in this case, then this shell will stall soon. However, 
the pressure may well be lower than the above value in this region,
since part of the shell is seen outside the disk. Other expanding 
shells (incomplete arclike shapes) with velocities possibly as low 
as 10 to 15 km s$^{-1}$ with radii $\sim$ 200 pc are also observed 
in the central region \citep{lok87}, suggesting a sound speed less 
than the 40 km s$^{-1}$ figure used above. 

(2) Observations of nearby bubbles in our own Galaxy and in the 
Magellanic Clouds indicate that the simple adiabatic bubble/shell 
theory \citep{wea77, mak87} coupled with the mechanical luminosities 
calculated by {\it Starburst99} for this paper leads to significant
overestimates of the bubble pressure and hence the shell radius 
\citep[e.g.][and references therein]{oag04}. Either the wind power 
is lower or some hot bubble gas escapes from the bubble interior. 
In addition, Dopita et al. (2005) argued that the conventional 
bubble/shell dynamical model may overestimate the winds and 
supernova mechanical power. Another argument is that gravitational 
instability may induce new star formation inside the shells. 
If such effects were present, they would have an impact on the 
estimate of the total stellar mass and luminosity in our model, 
as described in more detail in point (3).

(3) The bolometric luminosity for the best fit cluster mass of 
3.7 $\times$ 10$^6$ M$_{\odot}$ and best fit age of 5.6 Myr is 
1.5 $\times$ 10$^9$ L$_{\odot}$  (based on {\it Starburst99} model). 
The observed IR luminosity of M 82 disk is about 
3.0 $\times$ 10$^{10}$ L$_{\odot}$. Since the bolometric luminosity 
should be an upper bound to the IR luminosity from the same stars, 
the shortfall in the model luminosity is at least a factor of 20. 
This shortfall is similar to that (also about a factor of 20) 
between the star formation rate derived from our continuous star 
formation model ($\sim$ 0.5 M$_{\odot}$ yr$^{-1}$), and the measured 
star formation rate $\sim$ 5 - 10 M$_{\odot}$ yr$^{-1}$ for the 
center of M 82 \citep[e.g.][]{deg01, lap04}. Hence, the stellar 
cluster mass needed according to the model to produce the observed 
CO luminosity is not sufficient to account for all of the stellar 
luminosity or young stellar mass in M 82.

There are several reasons that our shell ensemble model may have 
overestimated the swept-up gas mass and the line emission for 
a given cluster mass, or equivalently underestimated the stellar 
mass and luminosity required for a given swept-up gas mass. 
Points (1) and (2) above show that our model itself may be 
fundamentally optimistic in its impact on the ISM, i.e. the 
model shells may be too big for the stellar mass  which generates 
them, thus leading to an overestimate in the swept-up shell masses 
and CO luminosity per unit stellar mass. The radii of the shells 
are larger than they would be in a more realistic model where 
the pressure of the ISM is included and where the effects of 
lower mechanical luminosity and leakage of bubble gas are included. 
These might be major effects and they both act in the same direction. 
If the shells at the best fit age are smaller, then we simply need 
more of them to build up the H$_2$ mass sufficient to explain 
the observed CO flux. In particular, if the shells were to stall 
early at radii about half that in our model, then a model which 
includes this effect would require about eight times the cluster 
mass for the shell ensemble. This factor already accounts for 
much of the missing stellar luminosity/mass. Thus the stellar 
mass required is very sensitive to the adopted model. 
Additional simulations to test a stalled-shell scenario will
be carried out in the near future. 

There are several other possible factors contributing to the 
shortfall in stellar luminosity represented by our model. 
Some SCs blow their shells out of the disk and are not detected, 
some or perhaps even most OB stars do not form in SSCs, and 
perhaps earlier generations of stars will augment the FIR 
luminosity to some degree.

(4) Our model neglects the emission from the low density ambient ISM 
($n_{ism}$ = 30 cm$^{-3}$), due to the lack of knowledge of the 
structure of this component in a starburst galaxy, and to the lack 
of direct observational data of this gas component that could be 
used to distinguish this gas and its physical state from the shell 
emitting gas in our models. If the ambient medium were uniform, 
as assumed in the model, it would produce no observable emission, 
since the density is too low to excite even the first excited 
rotational level. If, however, 
the ambient gas is assumed to be highly non-uniform, as is more likely 
the case, we can use the total mass adopted for the sampled volume 
to estimate its CO emission by simply using the $X$-factor for the 
CO emitting gas in our own Galaxy. We find that, before the shells 
form, the $^{12}$CO(1-0) emission from this ambient gas component 
would be about 54\% of the total current emission within the central 
1 kpc region. Note however that at high-$J$, there may be only very 
weak emission from this ISM component. Hence, the consequence of adding 
the emission of this lower density ambient gas component to the model 
would be to reduce the apparent excitation of the combined emission, 
especially the lower-$J$ transitions.

(5) A fixed microturbulent Doppler velocity ($\delta$$v_D$ = 1.5 km s$^{-1}$) 
is used in our model, as a heating mechanism in the PDRs and as a broadening 
mechanism for the molecular line profile. However, the choice of turbulent 
velocity directly affects the computed CO line flux densities. In the 
optically thick case, the flux is directly proportional to the turbulent 
velocity, whereas in the optically thin case there is no dependence. 
Since the lower-$J$ transitions are optically thick, more so than the higher 
level transitions, we anticipate that the use of a higher turbulent velocity 
would reduce the slope of the line SED (see Fig.~\ref{lsed_co1}). 

(6) The assumption of the semi-infinite plane-parallel geometry in our 
PDR code is certainly a limitation, as the FUV intensity may be
underestimated due to leakage of radiation from the region exterior 
to the cloud/shell, resulting in an increase in the local mean intensity
at the edge of the slab. More advanced models of the shell geometry 
are simply beyond the scope of the PDR code at the time, and solving 
for the radiation field from both sides of the slab dramatically increases 
the computation time and would make the code too slow to run.

(7) Other effects will invalidate our model for later stages of 
shell evolution. For example, after 30 Myr the largest radius of 
the shells in our model ensemble is about 678 pc. Thus, it will 
be merged with other shells, an effect which is not included 
in the model. It will also have extended beyond the scale height 
of the disk (300 pc along the minor axis), and be subjected to 
shear due to differential galactic rotation. However, these problems 
occur at ages older than our CO best fit model, and they should not 
significantly affect our best fit result.

Overall, our analysis shows that the sub-mm/mm line emission 
reflects the recent star formation history in a starburst galaxy.
The foregoing list of caveats and limitations ultimately limit 
the precision with which one can obtain a realistic age for the 
starburst by the method described in this paper. Some of the effects 
described lead to an underestimate and some to an overestimate. 
Thus, to some extent, the effects are liable to cancellation. 
However, the one factor upon which the ages depend most strongly 
is stellar evolution, since the evolutionary state of the cluster 
governs the flux of FUV emission incident upon the shell, and 
this in turn has an important influence on the SED of the molecular 
line emission. A consequence is that the age is unlikely to be 
profoundly affected by the effects listed. This also means that 
there should be little surprise with the agreement with other 
methods. However, the total cluster mass responsible is 
exceptionally sensitive to the model for the expansion of the shell, 
and consequently this quantity is less well determined than the age. 
In our model, it appears likely that this stellar mass in our 
cluster ensemble is severely underestimated.

\section{Conclusions} \label{sum}

This paper presents a {\it first} attempt at addressing the 
question of whether there is a {\it signal} in the FIR/sub-mm/mm 
molecular and atomic line data of the phase of a starburst. 
By comparing our evolving starburst models with available data 
of nearby starburst galaxies, notably M 82, we show that it is 
possible to (1) successfully model the time-dependent FIR/sub-mm/mm 
line emission of molecular and atomic gas; (2) relate the observed 
molecular line properties of a starburst galaxy to its age, and 
hence to constrain the global star formation history; and (3) examine 
the possible relevance to the formation of the molecular rings 
in M 82.

In essence, we have provided a {\it complementary} study to the 
previous work on estimating the age(s) of starburst in M 82 using 
quite different methods. In particular, the method is analogous to 
that of Efstathiou et al. (2000), which considered the observable 
effects of an evolving cluster on the IR emission from the 
surrounding expanding dust shell. We have also provided support 
for the {\it hypothesis} of molecular ring formation in the center 
of M 82.

The main conclusions drawn from comparisons of our ESbM model with 
the observation are:

1. There is good agreement between our supershell kinematic 
model and the observations of the expanding supershell centered 
around the presumed supernova remnant SNR 41.9 + 58 in M 82. 
The agreement supports the {\it hypothesis} that this supershell 
is created by strong winds from a young star cluster with a total 
mass of 2.2 x 10$^6$ M$_{\odot}$ which formed at its center 
about 0.8 Myr ago, and the total mechanical energy needed for 
the creation of this supershell is about 1.5 $\times$ 10$^{54}$ ergs.
This is the energy equivalent of the winds associated with 
$\sim$ 1700 O stars (each with $m_{\ast}$ $\ge$ 40 M$_{\odot}$). 
Our model also shows that there should be excess CO emission at 
high excitation transitions in this supershell. This is consistent 
with the provisional detection of such excess emission at 
$^{12}$CO(6-5) in the region of this supershell seen after the 
surrounding disk emission is removed. Both agreements suggest 
that the set of evolving starburst models we have put forward 
in this paper can be used to interpret other shells in M 82 or shells 
in other starburst galaxies.

2. The {\it age} derived from our analysis of CO line SEDs and line 
ratio diagrams using an instantaneous burst model is 5 - 6 Myr 
for the central 1 kpc region, with most of the CO emission arising 
from the central 680 pc region. The age derived from our atomic data 
is slightly older (10 Myr) for a larger area (i.e. 1,2 kpc)
We suggest that these two ages may be a result of sampling different
regions. It is unclear from our analysis whether they refer to the 
same period of star forming activity or to two spatially separated 
independent bursts. A more sophisticated model and more data are 
needed to clarify the picture. We do note however that our extended 
starburst model result also shows that a uniform star formation rate 
over the past 10 Myr can also produce a satisfactory model fit to 
the $^{12}$CO data. Hence, the star formation in M 82 can be either 
viewed as instantaneous burst occurred 5 - 6 Myr ago, or this epoch 
could represent a characteristic time about which recent star formation 
history is centered. These burst ages derived from our models are 
similar to the results found in optical and infrared studies. 
These results lead us to conclude that the observed FIR/sub-mm/mm 
line spectra of a starburst galaxy can be successfully modeled in 
terms of the evolutionary scheme of an GMC/shell ensemble, and such 
studies can usefully constrain the age(s) or star formation history 
of a starburst galaxy. 

The starburst ages derived from our model are dependent on a great 
variety of assumptions, e.g. the initial upper mass limit of the 
cluster spectrum. We find that the model can provide acceptable fits 
to the data only if the dominant initiating starburst clusters are 
massive, at least 5 $\times$ 10$^5$ M$_{\odot}$, corresponding to 
a GMC mass of 2 $\times$ 10$^6$ M$_{\odot}$. The uncertainty of the 
derived age is also affected by many other model assumptions, and 
the effects of varying these assumptions have not been examined. 
These include, for example, the effect of including the CO emission 
(especially low-$J$ transitions) from the lower density ambient ISM, 
the effect of including a higher cosmic-ray ionization rate, and the 
effect of increasing the shell microturbulent velocity. Some of these 
effects would lead to an underestimate and some to an overestimate 
of the age, and hence to some extent, these effects would be expected 
to cancel each other out. However, since the evolutionary state of 
the cluster governs the flux of FUV radiation incident upon the shell, 
and this in turn has a pronounced effect on the SED of the molecular 
line emission, the stellar evolution is a crucial factor in constraining
the derived age. The starburst stellar mass and luminosity predicted 
by our models are significantly underestimated, based on a comparison 
with the observed FIR luminosity which is a factor of about twenty 
larger than our model value for the total luminosities of the clusters. 
Probable causes for this underestimate include (1) the neglect of the 
effects of the pressure exerted by the ambient gas, resulting in an 
overestimate of the shell radii; and (2) an overestimate of the 
supernova mechanical power which would also lead to an overestimate 
of the shell radii. Including these effects would allow more stellar 
luminosity in the starburst for the mass of gas swept up in the shells 
required to match the CO data. Hence, the shortfall in our predicted 
stellar luminosity tends to support the widely held idea that bubbles/shells
grow more slowly than the simple bubble theory predicts \citep{wea77, mak87}. 

Our model also cannot provide a basis for incorporating higher density 
tracers (e.g. HCN and HCO$^+$), Because the critical densities are, for 
example, approximately 10$^5$ and 10$^7$ for the (1-0) and (4-3) transitions 
for both molecules. The HCO$^+$ intensities may be affected by cosmic-ray
ionization when it becomes the dominant heating source in the gas.
Both HCN and HCO$^+$ are associated with dense shells and their parent 
GMCs seen only at the earliest phase of the starburst evolution.
But another principal source is required, most probably dense gas associated 
with the cores of potential star forming regions, not included in our model. 
If the dense cloud core component was included in our model, the effect
of a higher cosmic-rate ionization rate (in M 82) should be taken into
account.

3. The results of the model analysis described above (item 2), also 
yield insights into the total gas content and its structure. For example, 
the total H$_2$ gas mass $\sim$ 2 - 3.4 $\times$ 10$^8$ M$_{\odot}$, 
is consistent with that measured independently in the center of M 82. 
The inference is that the neutral ISM and possibly the molecular ring 
in the center of M 82 are largely the products of evolving shells. 
However, our interpretation concerning the ring formation is not unique, 
and the rings may also be created by Linblad resonance instabilities 
associated with the gravitational effects of the bar.

\acknowledgments

I would like to thank my Ph.D. thesis advisor E. Seaquist for
his guidance and support. I also thank J. Yates at UCL for discussions 
on the {\it SMMOL}, T. Bell at Caltech and S. Viti at UCL for 
discussions on the {\it UCL\_PDR}, and C. Leitherer for discussions 
on the {\it Starburst99}. I am grateful to the Canadian Institute 
for Theoretical Astrophysics at University of Toronto for their 
generosity in allowing me to use their computing facility to run 
the PDR and Radiative Transfer simulations. I thank Space Telescope 
and Science Institute and National Radio Astronomy Observatory 
for kindly hosting me as their pre-doc visiting student. I also 
thank the referee for his careful reading, helpful comments, and 
his patience. This research was supported by a research grant from 
the Natural Sciences and Engineering Research Council of Canada to 
E. R. Seaquist, and a Reinhardt Graduate Student Travel Fellowship 
from the Department of Astronomy and Astrophysics at the University 
of Toronto.

\end{document}